\documentclass[]{aa}
\usepackage{graphicx}
\usepackage{dirtytalk}
\usepackage{currvita}
\usepackage[T1]{fontenc}
\usepackage{txfonts}
\usepackage{hyperref}
\usepackage{xcolor}
\hypersetup{
colorlinks,
linkcolor={red!80!black},
citecolor={blue!80!black},
urlcolor={blue!80!black}
}

\newenvironment{myfont}{\fontfamily{ppl}\selectfont}{\par}
\DeclareTextFontCommand{\textmyfont}{\myfont}

\newcommand{\code}[1]{\texttt{#1}}

\def\nifs{\iso{56}Ni}

\def\cm3{cm$^{-3}$}
\def\kms{\mbox{km~s$^{-1}$}}
\def\msunyr{$M_{\odot}$\,yr$^{-1}$}

\def\rsun{$R_{\odot}$}

\def\msun{$M_{\odot}$}

\def\one{\ts {\,\sc i}}
\def\two{\ts {\,\sc ii}}
\def\three{\ts {\,\sc iii}}
\def\four{\ts {\,\sc iv}}
\def\five{\ts {\sc v}}
\def\six{\ts {\sc vi}}
\def\beq{\begin{equation}}
\def\eeq{\end{equation}}

\def\lesssim{\mathrel{\hbox{\rlap{\hbox{\lower4pt\hbox{$\sim$}}}\hbox{$<$}}}}
\def\gtrsim{\mathrel{\hbox{\rlap{\hbox{\lower4pt\hbox{$\sim$}}}\hbox{$>$}}}}

\def\one{{\,\sc i}}
\def\two{{\,\sc ii}}
\def\three{{\,\sc iii}}
\def\four{{\,\sc iv}}
\def\five{{\sc v}}
\def\six{{\sc vi}}

\newcommand{\gcc}{\ensuremath{\rm{g\,cm}^{-3}}}

\def\cmfgen{{\code{CMFGEN}}}
\def\heracles{{\code{HERACLES}}}

\def\ergs{erg\,s$^{-1}$}

\newcommand{\iso}[2]{\ensuremath{^{#1}\rm{#2}}}

\begin{document}

   \title{Using spectral modeling to break light-curve degeneracies of type II supernovae interacting with circumstellar material}

   \titlerunning{Type II-P SNe interacting with CSM}

\author{
Luc Dessart\inst{\ref{inst1}}
\and
W. V. Jacobson-Gal\'an\inst{\ref{inst2}}
 }

\institute{
    Institut d'Astrophysique de Paris, CNRS-Sorbonne Universit\'e, 98 bis boulevard Arago, F-75014 Paris, France\label{inst1}
   \and
Department of Astronomy and Astrophysics, University of California, Berkeley, CA 94720, USA\label{inst2}
}

   \date{}

  \abstract{A large fraction of red-supergiant stars seem to be enshrouded by circumstellar material (CSM) at the time of explosion. Relative to explosions in a vacuum, this CSM causes both a luminosity boost at early times as well as the presence of symmetric emission lines with a narrow core and electron-scattering wings typical of type IIn supernovae (SNe). For this study, we performed radiation-hydrodynamics and radiative transfer calculations for a variety of CSM configurations (i.e., compact, extended, and detached) and documented the resulting ejecta and radiation properties. We find that models with a dense, compact, and massive CSM on the order of 0.5\,\msun\ can match the early luminosity boost of type II-P SNe but fail to produce type IIn-like spectral signatures (also known as ``flash features''). These only arise if the photon mean free path in the CSM is large enough (i.e, if the density is low enough) to allow for a radiative precursor through a long-lived (i.e., a day to a week), radially extended unshocked optically thick CSM. The greater radiative losses and kinetic-energy extraction in this case boost the luminosity even for modest CSM masses -- this boost comes with a delay for a detached CSM. The inadequate assumption of high CSM density, in which the shock travels essentially adiabatically, overestimates the CSM mass and associated mass-loss rate. Our simulations also indicate that type IIn-like spectral signatures last as long as there is optically-thick unshocked CSM. Constraining the CSM structure therefore requires a combination of light curves and spectra, rather than photometry alone. We emphasize that for a given total energy, the radiation excess fostered by the presence of CSM comes at the expense of kinetic energy, as evidenced by the disappearance of the fastest ejecta material and the accumulation of mass in a dense shell. Both effects can be constrained from spectra well after the interaction phase.
}


   \maketitle


\section{Introduction}

In recent years, the observation of narrow, electron-scattering broadened, rather than broad, Doppler-broadened lines in the earliest spectra of some hydrogen-rich, type II supernovae (SNe) has exacerbated the interest in the final stages of massive star evolution and the potential for intense mass loss in the years or decades leading  to core collapse. One emblematic object that increased this interest was SN\,2013fs\ \citep{yaron_13fs_17}, which exhibited symmetric profiles with a narrow core and extended wings in lines of highly ionized species. Such signatures, which are typical of interacting SNe (generally known as type IIn SNe; \citealt{schlegel_iin_90}), persisted in all spectra of SN\,2013fs taken within about 20 hours after shock breakout and this required some circumstellar material (CSM)  in the direct vicinity of the progenitor red-supergiant (RSG) star \citep{yaron_13fs_17}.\footnote{In this paper, to refer to these spectral signatures, we use the term type IIn-like since their appearance and formation process is the same as those observed in bona fide interacting, type IIn SNe. We will not use the more popular term flash features since the supply of ionizing photons starts with shock breakout but continues for days in typical type II SNe. In general, the disappearance of these features does not result from a shortage of ionizing photons but from the progressive reduction of slow dense CSM ahead of the shock.} The mass of this CSM has been inferred to range from 0.001\,\msun\ \citep{yaron_13fs_17}, to 0.01--0.1\,\msun\ \citep{d17_13fs}, up to 0.5\,\msun\ \citep{morozova_2l_2p_17,moriya_13fs_17}. Furthermore, this CSM has been estimated to extend from about 1 to $5 \times 10^{14}$\,cm depending on whether the modeling focused only on the light curve \citep{morozova_2l_2p_17} or whether it used spectroscopic constraints (i.e., the duration over which the narrow spectral features persist; \citealt{yaron_13fs_17}). Hence, while there is a broad consensus on the fact that some CSM must be present at the surface of the SN\,2013fs progenitor at the time of explosion, there is poor consensus on its properties.

Since SN\,2013fs and the publication of \citet{yaron_13fs_17}, additional observations have been obtained and new models have been produced. When considered as a whole, the majority of type II-P SNe tend to show signs of an interaction at early times as evidenced through the transitory presence of narrow symmetric H\one\ or He\two\ lines  \citep{bruch_csm_21,bruch_csm_22} or their photometric rise times \citep{forster_csm_18}. While pre-SN activity is recorded in most SNe classified as type IIn \citep{khazov_flash_16,strotjohann_presn_21}, it may also occur in objects that eventually appear as standard type II-P SNe. SN\,2020tlf is one such example with an extraordinary combination of properties including a 120\,d long pre-SN activity with a bolometric luminosity of 10$^{40}$\,\ergs\ until explosion, followed by a week-long phase of interaction analogous to what was observed in the type IIn SN\,1998S \citep{leonard_98S_00}, transitioning after a few weeks into a normal type II-P SN, and eventually entering its nebular phase powered by \nifs\ decay with signatures typical of a relatively light, 12\,\msun\ progenitor star \citep{wynn_20tlf_22}. Such a prolonged interaction phase following shock emergence requires a dense and extended CSM, more extended than inferred in SN\,2013fs.

Thereotically, these events have been modeled in the context of a compact, dense, and massive CSM directly at the surface of the progenitor star \citep{morozova_2l_2p_17}, but such CSM properties are in tension with the sometimes prolonged survival of type IIn signatures seen in these events. For example, for the case of SN\,2020tlf or 1998S, the type IIn signatures persist for at least a week and cannot be accommodated for by dense material within a few $R_\star$. There thus seems to be at least two families of events (or CSM configurations) in the current sample of type II-P SNe with early-time signatures of an interaction, with either short-lived ($<1$\,d) or week-long type IIn signatures, hence associated with compact or extended CSM. Each family may be produced by a distinct mechanism.

The most widely-accepted mechanism is a CSM that forms from a super-wind phase \citep{quataert_shiode_12,fuller_rsg_17} taking place in the months to years before core collapse \citep{morozova_2l_2p_17,moriya_13fs_17,d17_13fs,morozova_sn2p_18}. The ambiguity of the observables opens alternative possibilities, such as a progressive mass overloading taking place over the entire RSG lifetime, as proposed by \citet{d17_13fs} or more recently \citet{soker_csm_21}, or RSG envelope convection and associated instabilities \citep{kozyreva_21yja_22,goldberg_3d_rsg_22}.

In this paper, we investigate the radiative properties of type II-P SNe with early-time signatures of an interaction with the specific aim of understanding what lies behind the scatter in CSM properties inferred from different studies. In the next section, we present the various CSM configurations that we used for this work. Then, in Sect.~\ref{sect_degen}, we highlight the degeneracy from light-curve modeling and in particular how different CSM properties can yield essentially the same bolometric light curve but different dynamical and spectral evolution. Section~\ref{sect_grid} discusses the correlations between ejecta, photometric, and spectroscopic properties for a representative grid of CSM or wind mass-loss rates attached to the exploding RSG star. Section~\ref{sect_det} explores the impact of a CSM that is detached rather than touching the progenitor surface. In Sect.~\ref{sect_conc}, we discuss our results in a broader context and give our conclusions. The simulations presented in this work will be confronted to a large set of observations in a forthcoming paper (Jacobson-Gal{\'a}n et~al., in preparation).

\section{Numerical setup}
\label{sect_setup}

In this work, we perform radiation hydrodynamics calculations with the code \heracles\ (\citealt{gonzalez_heracles_07}; \citealt{vaytet_mg_11}; \citealt{D15_2n}; Sect.~\ref{sect_rhd}) and nonlocal thermodynamic equilibrium (NLTE) radiative transfer calculations with the code \cmfgen\ (\citealt{HD12}; \citealt{D15_2n}; Sect.~\ref{sect_rt}) for a variety of ejecta-CSM interactions in order to compute the evolution of the gas and radiation over time. We first start in the next section by describing  the initial conditions for the ejecta and CSM that are used for the \heracles\ calculations. The \cmfgen\ calculations are done as a post-processing step on selected snapshots from the \heracles\ calculations. The overall approach is similar to that presented in \citet{d17_13fs} and more recently in \citet{wynn_20tlf_22}, to which the reader is referred for additional details and results for different ejecta-CSM configurations.

All calculations in this work assume spherical symmetry. While there is observational evidence for asymmetry from spectropolarimetry or line profile morphology in some interacting SNe (see, e.g., \citealt{leonard_98S_00} or \citealt{smith_ptf11iqb_15}), the presence of narrow, type IIn-like spectral signatures also indicates an optically-thick CSM that covers all sight lines to the SN so that the fast-moving material is obscured in all directions to earth (otherwise we would observe both Doppler-broadened lines and electron-scattering broadened lines). We thus make the ansatz here that asymmetry introduces moderate and localized variations of a typically large CSM density. In some rare cases, the CSM may reside in a disk or some highly asymmetric structure, and the assumption of spherical symmetry would be inadequate in this case but there is no observational evidence reported in the current literature for such highly asymmetric CSM in SNe IIP-CSM. The inferences made for extraordinary interacting SNe like 2009ip \citep{mauerhan_pol_09ip_14} or 2014C \citep{margutti_14C_16} should not be generalized to the CSM routinely observed around SNe II \citep{bruch_csm_21}. Obviously, the current simulations in spherical symmetry are a good starting point to investigate the impact of asymmetry on observables (see, for example, \citealt{D15_2n}; \citealt{vlasis_2n_16}; \citealt{kurfurst__sn_csm_20}).

\subsection{Initial conditions for the ejecta and the CSM}
\label{sect_init}

For the progenitor, we used model m15mlt3 from \citet{d13_sn2p}, as in the previous and similar study of \citet{d17_13fs}. Model m15mlt3 corresponds to a solar-metallicity 15\,\msun\ progenitor star dying as a RSG star with a surface radius  $R_\star$ of 501\,\rsun, a final mass of 14.08\,\msun, an H-rich envelope mass of 10.16\,\msun, and exploded to yield an ejecta of 12.52\,\msun, an asymptotic ejecta kinetic energy of $1.34 \times 10^{51}$\,erg, and 0.086\,\msun\ of \nifs. As in \citet{d17_13fs}, this model was taken just prior to shock breakout to avoid doing a time-consuming simulation of the whole explosion phase with \heracles.  Since the current study is mostly conceptual, using just one representative RSG progenitor and its associated standard type II-P SN model is sufficient. Varying the progenitor mass, the metallicity, the explosion energy etc would merely introduce moderate variations in the results but all conclusions presented here would hold at a qualitative level. The main driver controlling the properties of type IIP SNe interacting with CSM is the CSM -- this arises in part because the power released in this interaction swamps alternate sources, modifies drastically the spectral appearance, boosts the luminosity etc. In a future study, we will consider a much broader range of configurations, allowing the properties of both the progenitor, the explosion, and the CSM to vary.

\begin{figure}
\centering
\includegraphics[width=\hsize]{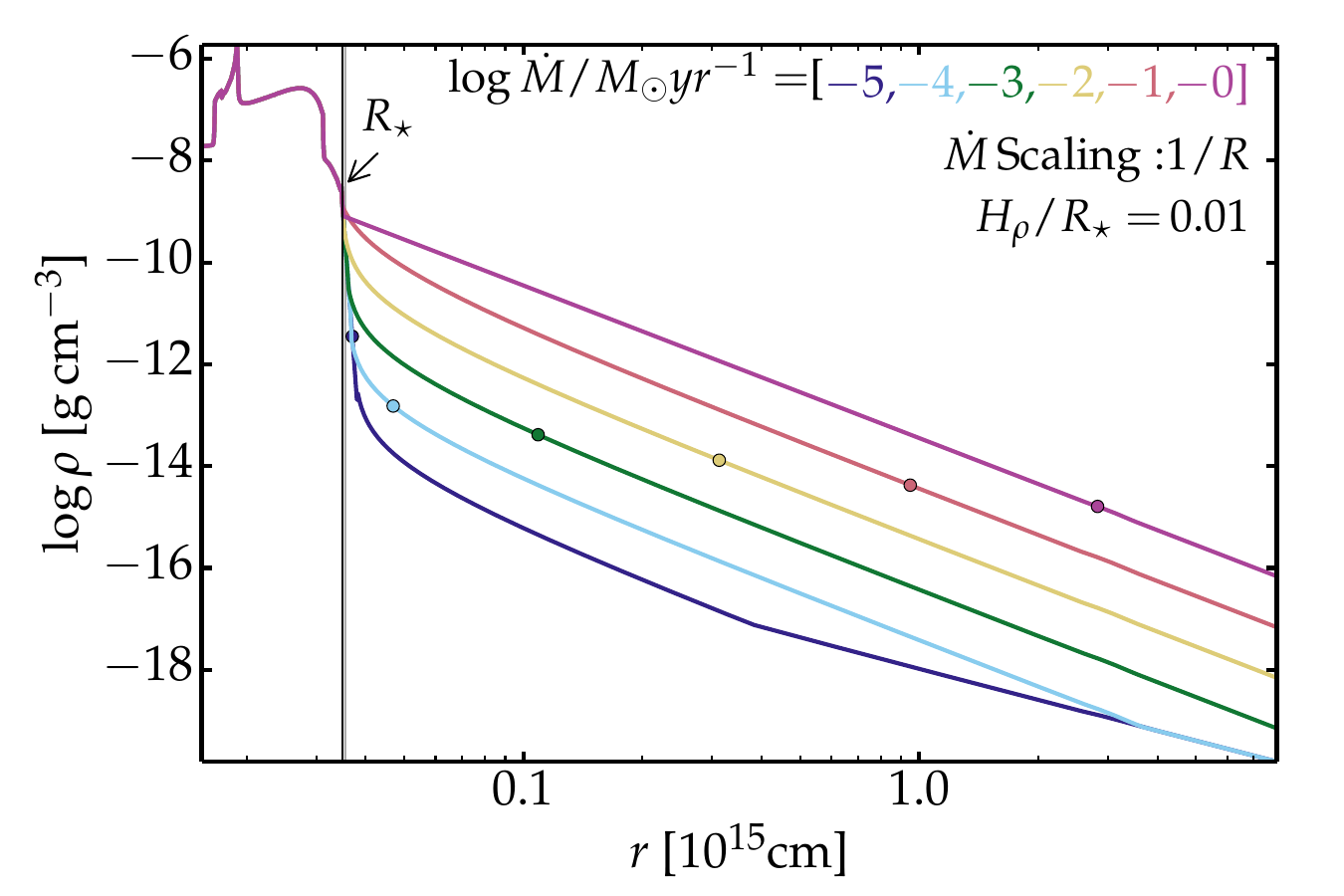}
\includegraphics[width=\hsize]{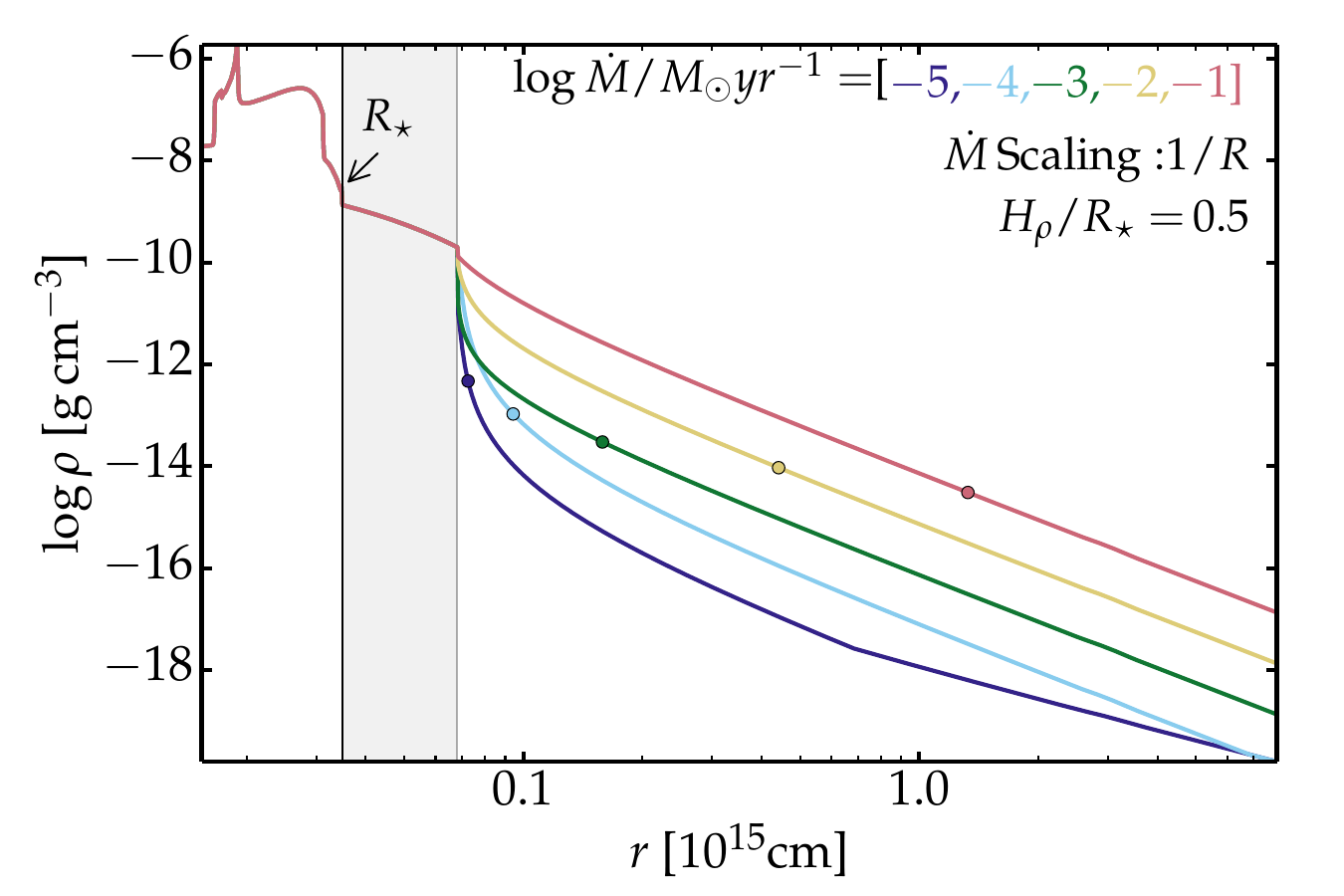}
\includegraphics[width=\hsize]{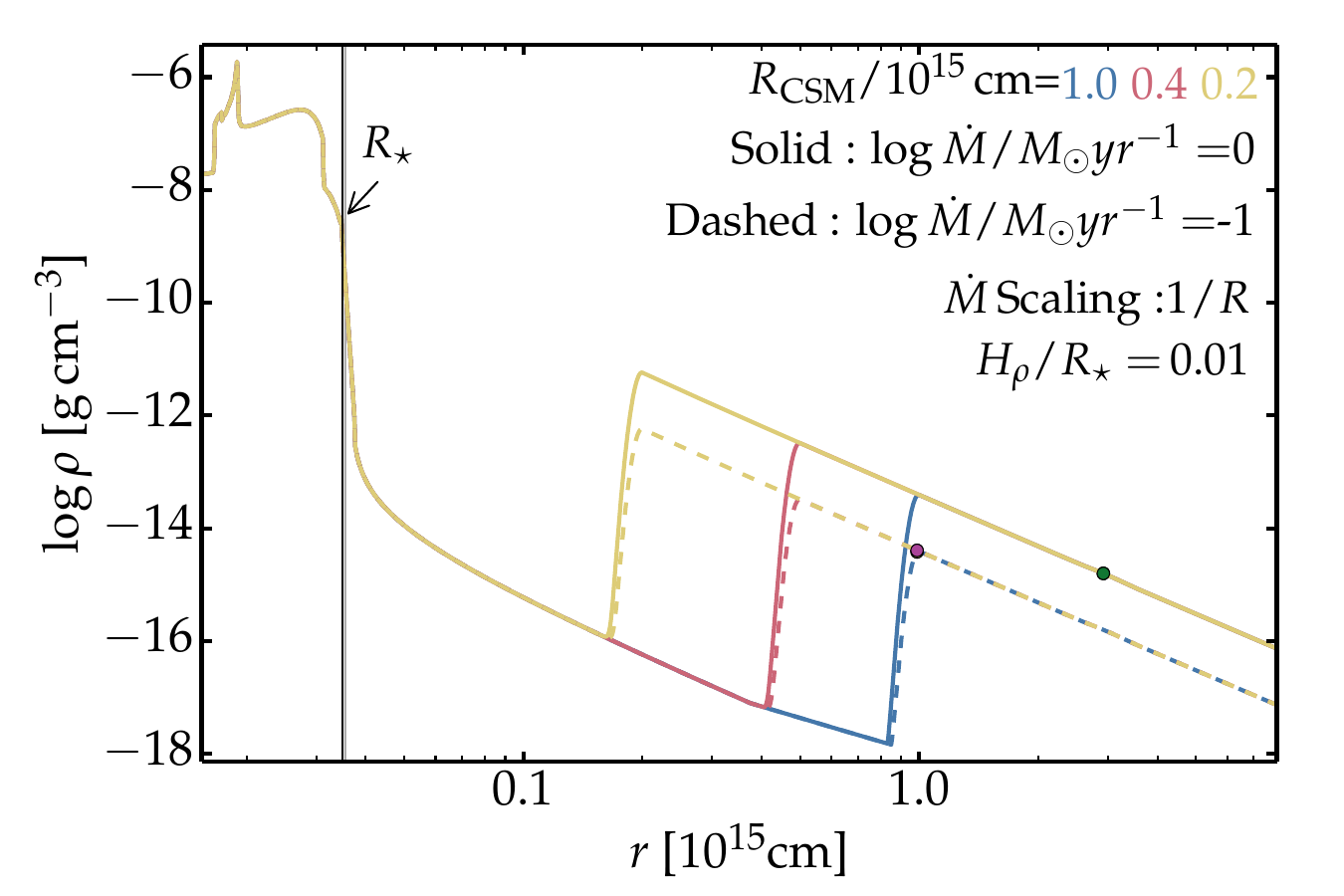}
\caption{Initial density structures corresponds to Case 1 (top), Case 2 (middle) and Case 3 (bottom). In each panel, we indicate the location of the stellar photosphere (arrow with $R_\star$ label and vertical line) of the original stellar model. The shaded area for case (2) corresponds to the dense CSM characterized by a large density scale height before transitioning to a wind solution. The dot marks the location of the photosphere assuming the material is ionized ($\kappa=$\,0.34\,cm$^2$\,g$^{-1}$). For details, see Section~\ref{sect_init}.
\label{fig_init}
}
\end{figure}

Beyond $R_\star$, we surrounded this prebreakout explosion model with some CSM. We explore three types of CSM, which reflect the various types that are routinely used in the community or may be expected from mass-losing RSG stars. These three cases, together with the sections where they are discussed, are the following (see also Fig.~\ref{fig_init}):
\begin{enumerate}
\item Standard wind from low to high mass-loss rates (Sects.~\ref{sect_degen}--\ref{sect_grid}).
\item Super-dense confined CSM (Sect.~\ref{sect_degen}).
\item Standard wind but detached from the progenitor surface (Sect.~\ref{sect_det}).
\end{enumerate}

Case (1) corresponds to a standard RSG star embedded in a steady-state wind whose strength covers from $10^{-5}$ up to 1\,\msunyr. For simplicity we adopted a constant CSM velocity of 50\,\kms. We ensured a smooth transition between the outermost layer of the progenitor at about 10$^{-9}$\,\gcc\ and the wind base (where the density is on the order of 10$^{-14}$ to 10$^{-12}$ \,\gcc) by inserting an ``atmosphere" with a density scale height $H_\rho$ of 0.01\,$R_\star$. These density structures are analogous to those used in \citet{d17_13fs} apart from the density break set at about $5 \times 10^{14}$\,cm, which was used in that former study to match the observed properties of SN\,2013fs.

Case (2) corresponds to a CSM structure similar to that used by \citet{morozova_2l_2p_17}, that is the CSM is very dense and confined to the immediate vicinity of the progenitor surface. Here, we achieved this by adopting an atmosphere with $H_\rho$ of 0.5\,$R_\star$ until a radius of about 1000\,\rsun\ where the density has dropped to $2 \times 10^{-10}$ \,\gcc. Beyond that external radius, we added a wind corresponding to various mass-loss rates, as in Case (1). This was required practically because \heracles\ is a Eulerian code and thus this outer region corresponds to the space into which the explosion will expand.  Strictly speaking, our density structures are analogous to those of \citet{morozova_2l_2p_17} for the case in which we employed $H_\rho$ of 0.5\,$R_\star$ combined with an external tenuous wind of 10$^{-6}$\,\msunyr. Such an abrupt change in density at the edge of the compact CSM does not seem physically consistent with what one may expect from a stellar eruption or super-wind event (e.g., a build-up in luminosity would progressively enhance the mass-loss rate leading to an extended distribution of material beyond that dense CSM).

Case (3) is similar to Case (1) except that we detached the CSM from the progenitor surface. This configuration was considered by \citet{moriya_maeda_12}. This case excludes the possibility of material stagnating in the vicinity of the progenitor since it must be driven off to sizable radii. This gap between the CSM and progenitor surface indicates a delay between the phase of high mass loss and the explosion, and the corresponding delay depends on the CSM velocity. We explored mass-loss rates of 0.1 and 1\,\msunyr\ and inner CSM radii of 2, 5, and 10 $\times 10^{14}$\,cm.

In all three cases, the CSM refers to the material above $R_\star$ and is composed of an atmosphere with a given scale height $H_\rho$ followed by a wind density structure (nominal $1/R^2$ density dependence, with an additional scaling -- see below). The distinction between atmosphere and wind is not really physical.  Here, it is used to differentiate the transition region between $R_\star$ and the wind. The atmosphere density profile goes as $\exp(-(R-R_\star)/H_\rho)$ and is considered at rest, and thus differs from the wind structure. When quoting the CSM mass, we mean the total mass of the atmosphere and wind above $R_\star$.

We ignore any acceleration region in the wind, from the hydrostatic base until large distances where the terminal velocity is reached. In reality, this acceleration region is complex in RSG stars, in particular because the low density winds of even standard RSGs may not be in steady state. One notorious example is the time-dependent aspherical mass distribution, made of upflows and downflows, observed in Betelgeuse  \citep{kervella_betelgeuse_09, kervella_betelgeuse_11,ohnaka_betelgeuse_11}.\footnote{\citet{moriya_13fs_17} explored the influence of this acceleration region. The corresponding wind density can be large and yield a large CSM mass if one adopts very low velocities of a few \kms. But then, such velocities make little sense physically since they are smaller than the convective and turbulent velocities at the RSG surface (see, for example, \citealt{goldberg_3d_rsg_22}).} Hence, the results here should be interpreted in terms of wind density rather than mass-loss rate.

In order to prevent a large wind density at large distances, which would tend to make a superluminous type IIn event for the high mass-loss rate cases, we enforced a scaling on the wind density by an extra factor of $1/R^s$ where $s$ is 1.0 or 0.5. This is mostly for numerical convenience to keep the CSM mass moderate and ensure an optically-thin outer boundary in \heracles. In all cases,  we adopted a density floor corresponding to a representative RSG wind mass-loss rate of  10$^{-6}$\,\msunyr. With this setup, the density smoothly drops until that density floor.

In all cases, the CSM was given a low temperature of 2000\,K initially. For the CSM composition, we adopted the corresponding values at the RSG surface of model m15mlt3 but limited to the five most abundant species at selected depth, namely H, He, O, Si, and Fe (i.e., \nifs\ prior to decay). \heracles\ then tracks these five species using passive scalars as part of the hydrodynamics.

\subsection{Radiation hydrodynamics simulations}
\label{sect_rhd}

The interaction configurations described above are used as initial conditions for 1D multigroup radiation-hydrodynamics simulations with the code \heracles. We use eight groups that cover from the ultraviolet to the far infrared: one group for the entire Lyman continuum, two groups for the Balmer continuum, two for the Paschen continuum, and three groups for the Brackett continuum and beyond. We adopt a simple equation of state that treats the gas as ideal with an adiabatic index of $\gamma=$\,5/3. We account for electron-scattering as well as bound-free opacity, which dominate over line opacity in ionized type II SN ejecta. The effect of line opacity is treated accurately in the post-processing step with \cmfgen.  In \heracles, the opacities are stored in a table with entries for density and temperature, with a corresponding ionization computed with a Saha solver \citep{D15_2n}. Five tables are used to cover from the H-rich outer regions to the metal-rich inner regions, and we interpolate in between according to the local mean atomic weight. The main focus here is on the early times when the spectrum formation region is located in the outer, H-rich layers, so most of the observables discussed here are influenced by the opacity and emissivity of H-rich material, hence associated with just one opacity table.

To follow the expansion of the preshock breakout ejecta and CSM until late times, we use a uniform grid out to $5 \times 10^{14}$\,cm and a logarithmic grid beyond and out to the maximum radius at $8 \times 10^{15}$\,cm. We use a total of 17408 grid points, which ensures a good resolution throughout the evolution of the interaction. To facilitate the comparison, we use the same radial grid in all simulations.

\subsection{Post-treatment with radiative transfer simulations}
\label{sect_rt}

At selected times in the \heracles\ simulations, we computed NLTE spectra with \cmfgen. We imported the properties of the gas (radius $R$, velocity $V$, temperature $T$, density $\rho$, and composition) and solved the radiative transfer using the nonmonotonic velocity solver and assuming steady state \citep{D15_2n}. The temperature was held fixed during the simulation because at the times considered, the temperature is controlled by the hydrodynamics and the power released by the shock in the optically-thick medium. Nonetheless, at fixed $T$, \cmfgen\ converges to a NLTE ionization (number density for all atoms and ions) that departs from the Saha, LTE solution. This approach works well as long as the continuum optical depth is at least of a few. When conditions turn optically thin, the current approach is inadequate and one must use an alternative approach (see, for example,  \citet{dessart_csm_22}  for SNe II interacting with an optically-thin wind or \citet{dessart_ibn_22} for SNe Ibn at late times).

We computed spectra at relatively early times, when the spectrum forms in the H-rich regions of the ejecta or CSM. The composition is therefore uniform and we used $X_{\rm He}=$\, 0.34, $X_{\rm C}=$\,0.00128,  $X_{\rm N}=$\,0.00329, and $X_{\rm O}=$\,0.00467 (and other metals at their solar metallicity value; the H mass fraction is obtained by requiring a normalization to unity). At early times, we included H\one, He\one-\two, C\one-\four, N\one-\four, O\one-\four, Mg\one-\two, Si\two, S\two, Ca\two, Cr\two-\four, Fe\one-\four, Co\two-\four, and Ni\two-\four. At later times, we dropped the high ionization stages and add the atoms or ions Na\one, Mg\one, Si\one, S\one, Ca\one, Sc\one-\three, and Ti\two-\three.

\begin{figure*}
\centering
\includegraphics[width=0.4\hsize]{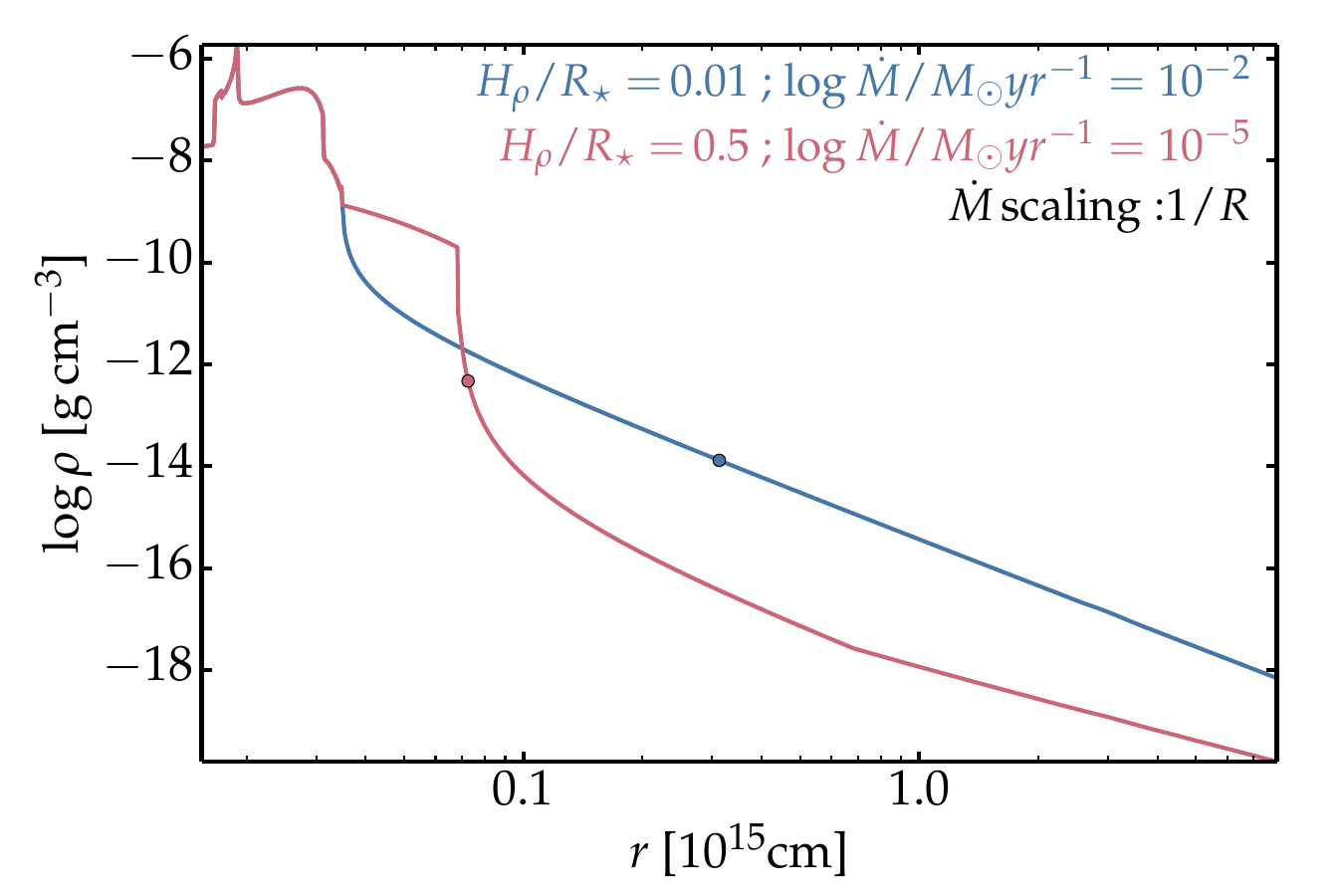}
\includegraphics[width=0.4\hsize]{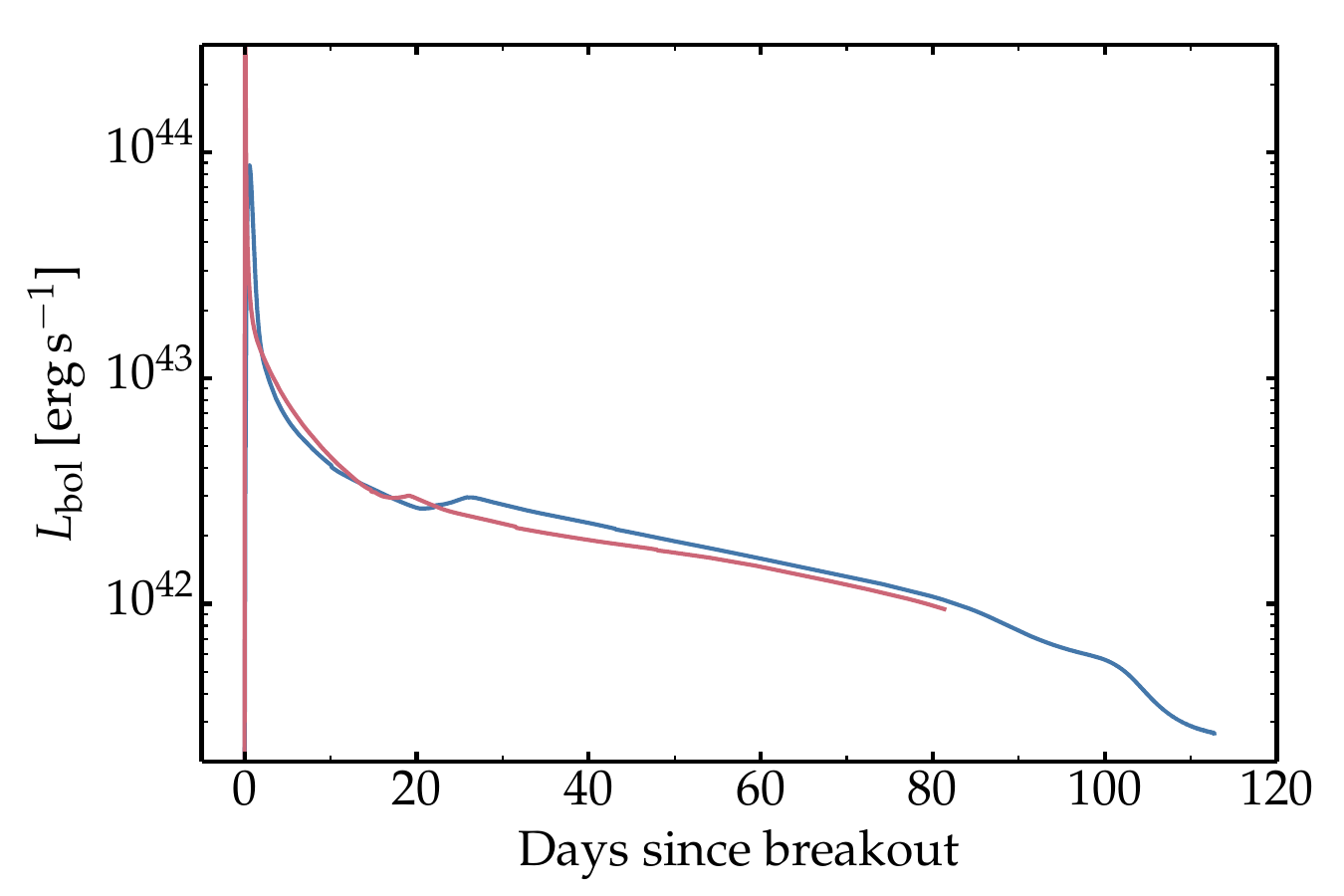}
\includegraphics[width=0.4\hsize]{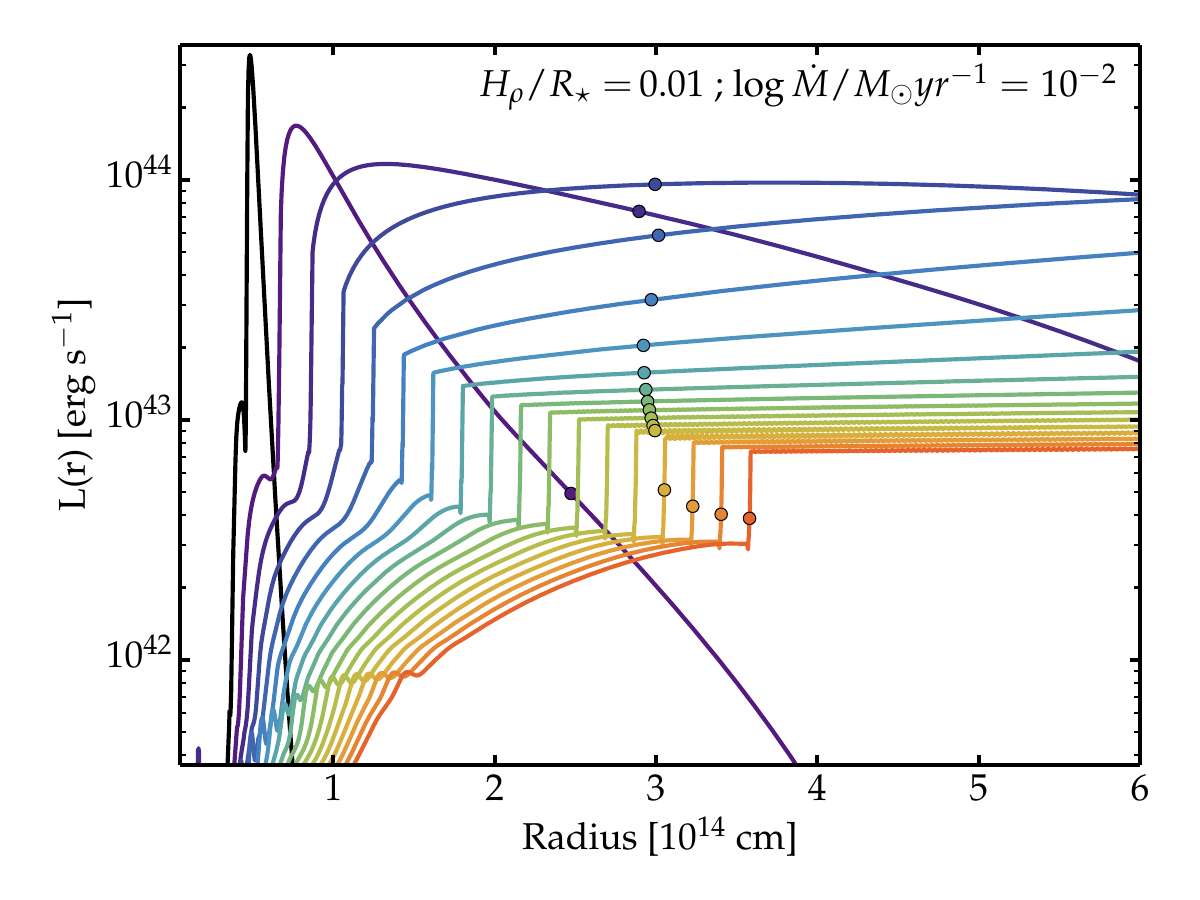}
\includegraphics[width=0.4\hsize]{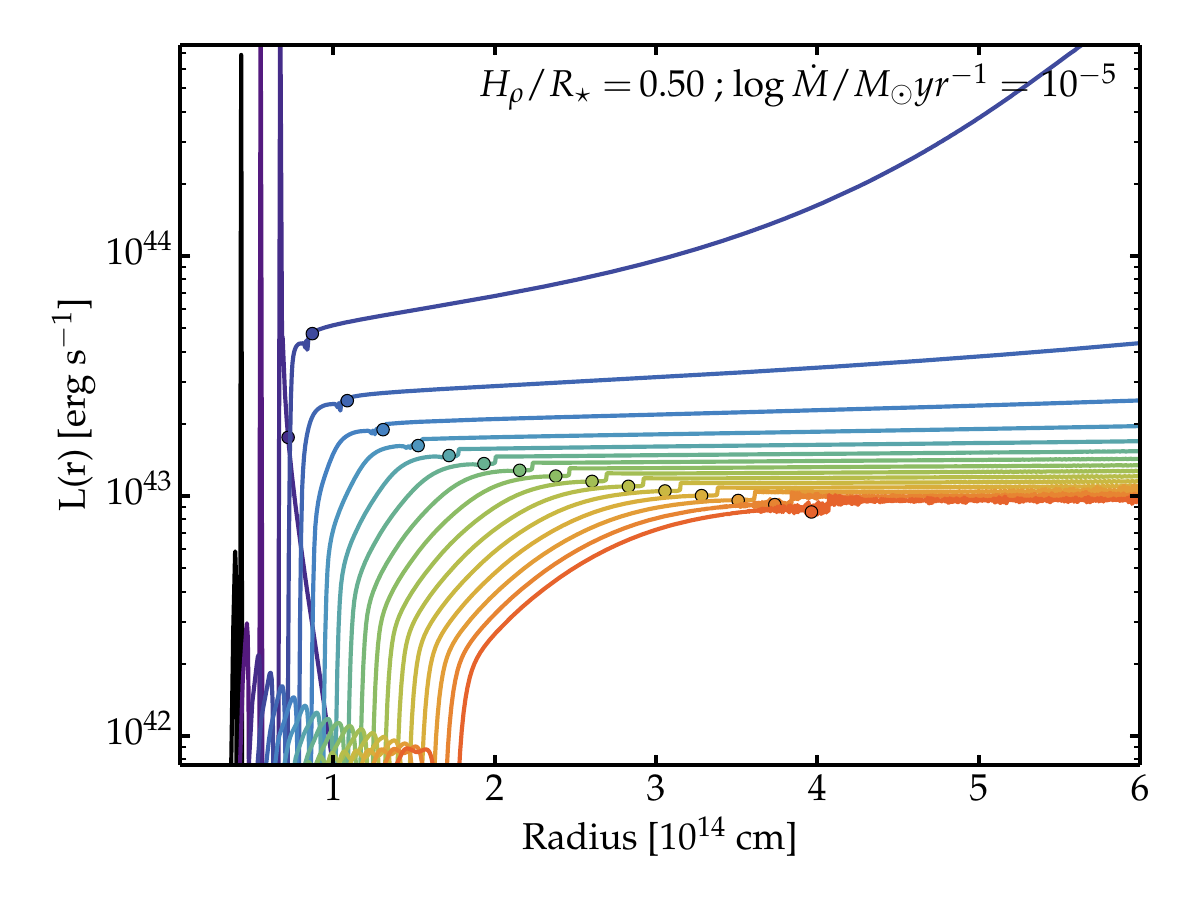}
\includegraphics[width=0.4\hsize]{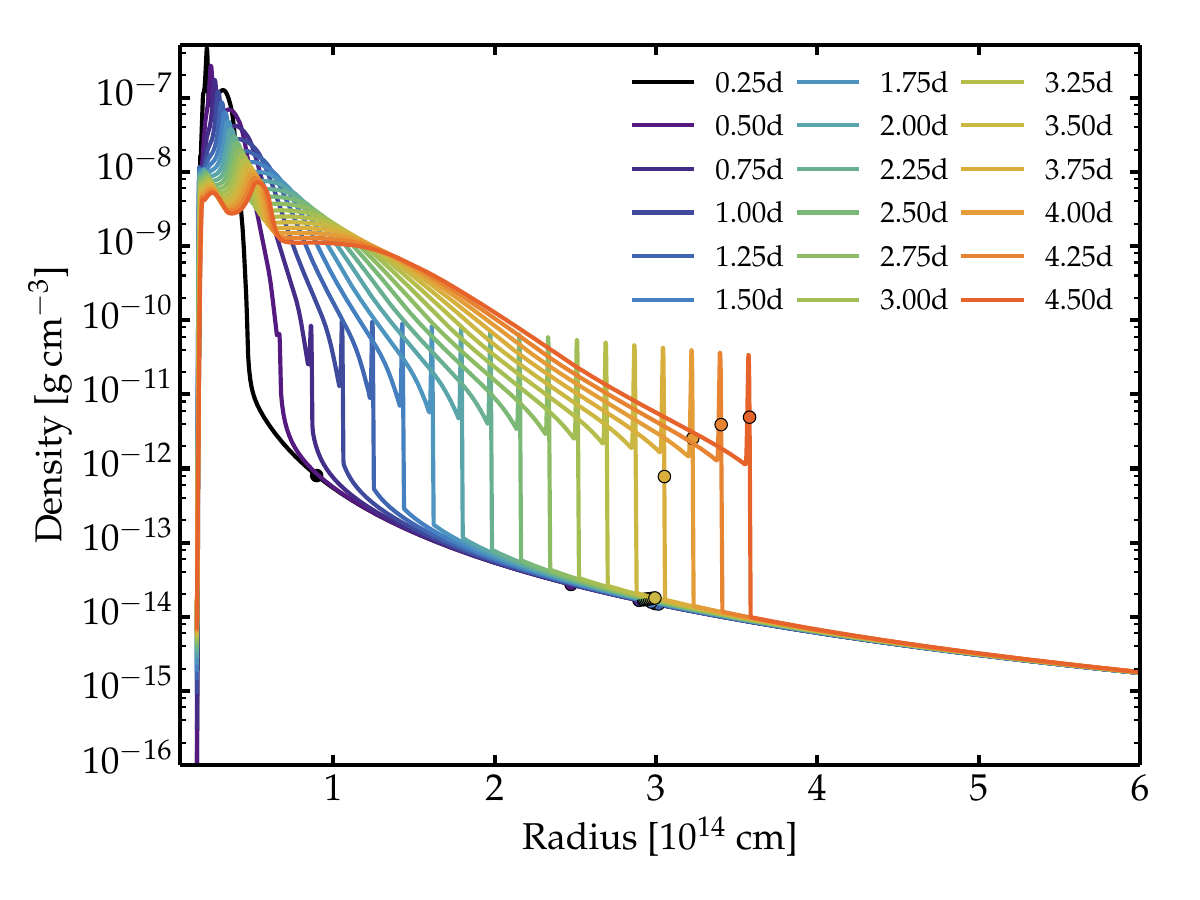}
\includegraphics[width=0.4\hsize]{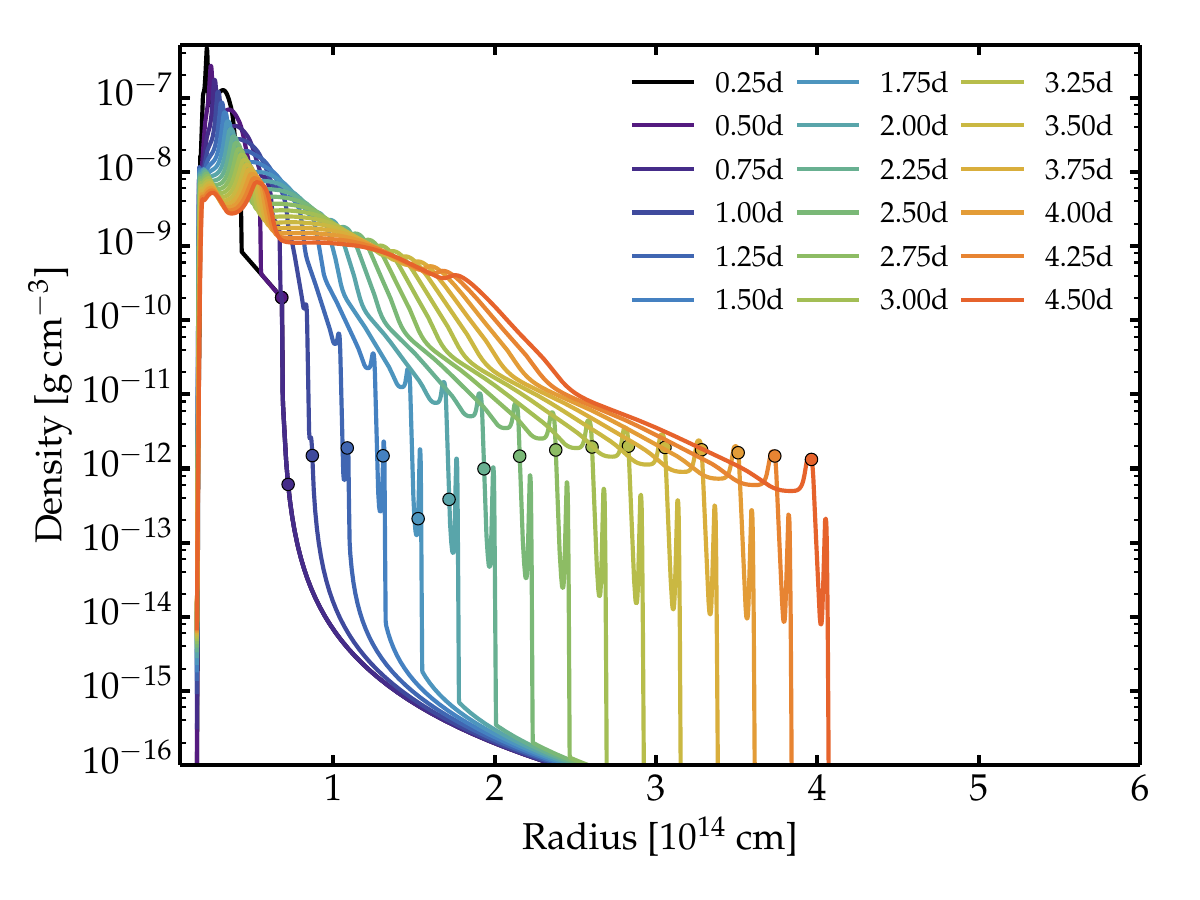}
\includegraphics[width=0.4\hsize]{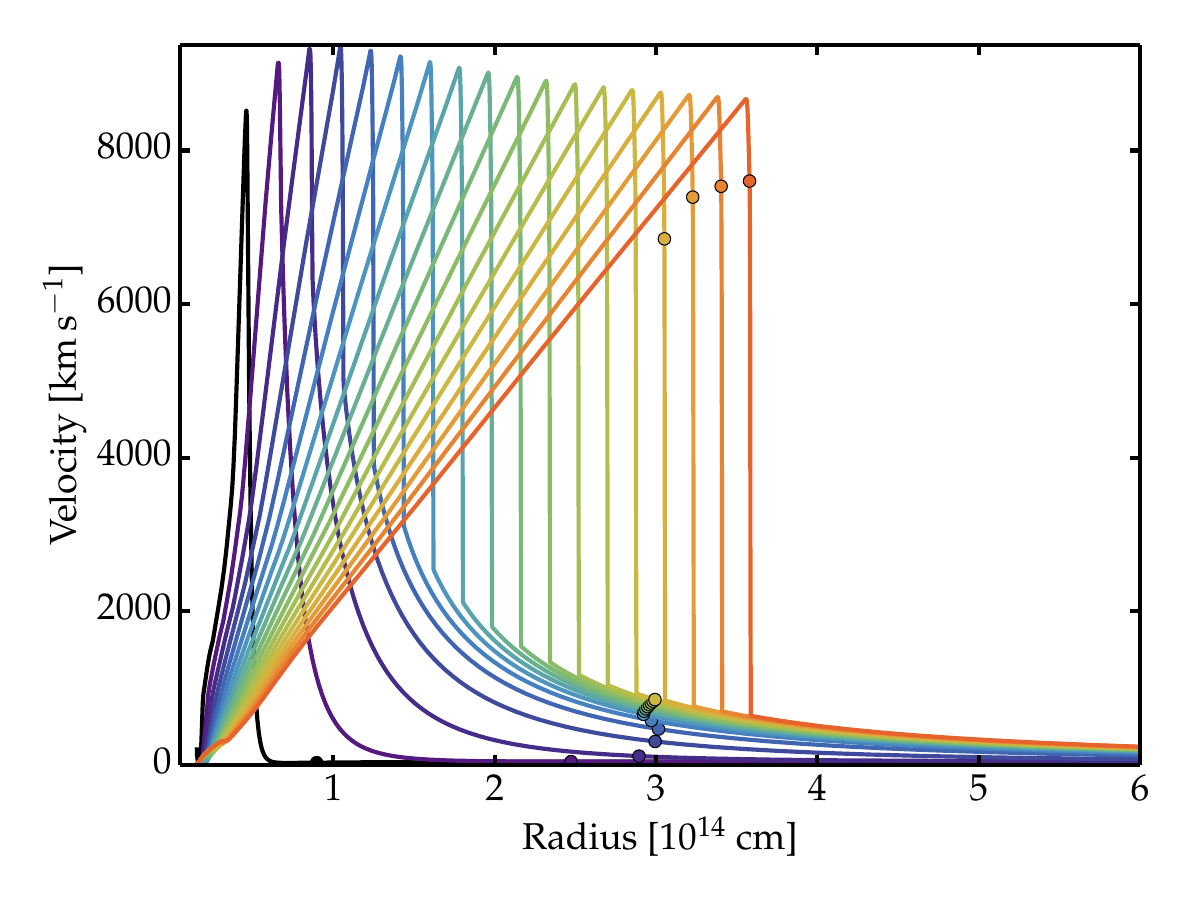}
\includegraphics[width=0.4\hsize]{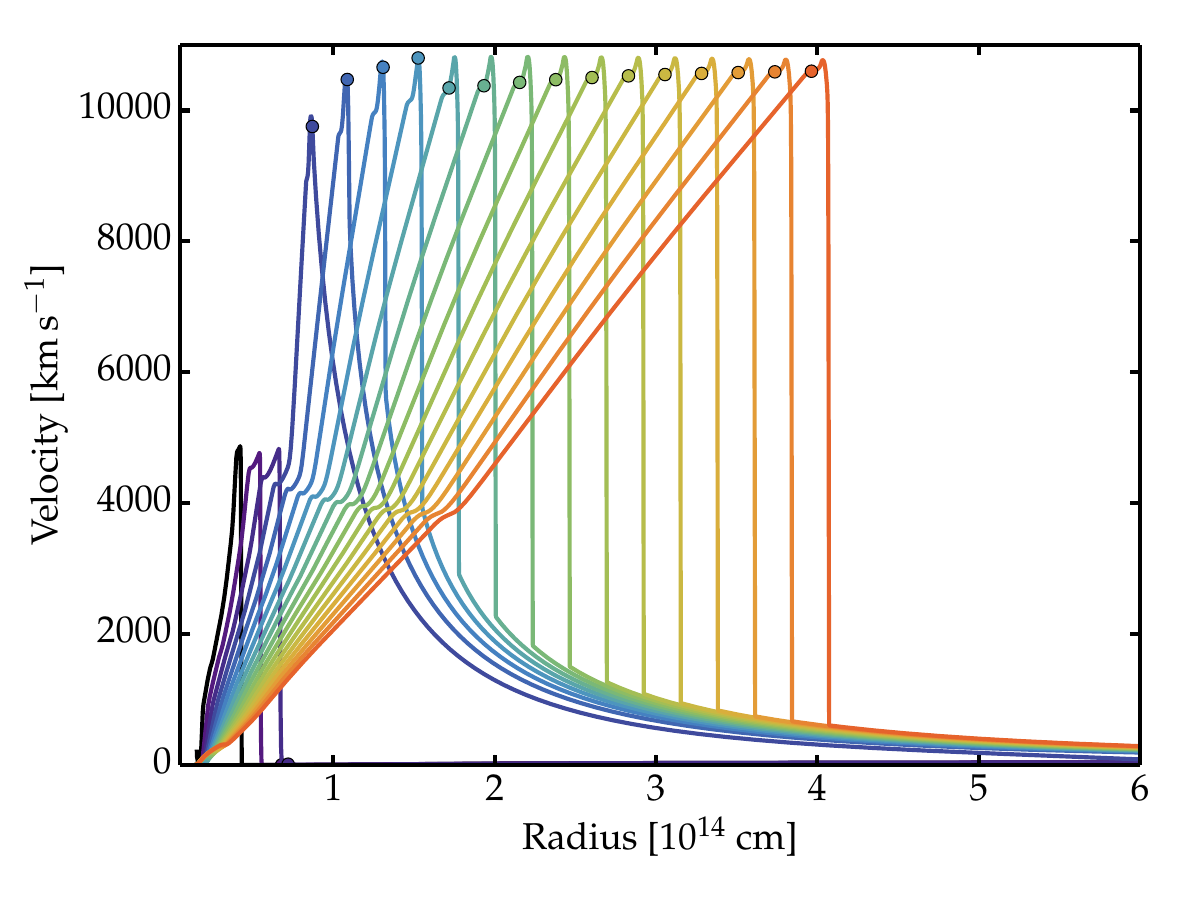}
\caption{Illustration of the light-curve degeneracy resulting from the interaction of ejecta with a CSM of 0.024 and 0.28\,\msun. The initial conditions for the interaction are shown at top left: case (1) corresponds to a model with a steep atmosphere followed by a dense wind and case (2) corresponds to a model with a shallow atmosphere followed by a tenuous wind. The resulting bolometric light curves from \heracles\ are shown at top right. The following three rows then show the luminosity, the density, and the velocity versus radius for first case (left column; 0.024\,\msun\ of CSM) and the second case (right column; 0.28\,\msun\ of CSM). The dot shows the location of the photosphere as computed in \heracles.
\label{fig_couple}
}
\end{figure*}

\section{Degeneracy of early-time light curves from RSG with CSM}
\label{sect_degen}

Before discussing the grid of simulations that we performed with \heracles, we first explore the degeneracy of bolometric light curves in SN simulations. This degeneracy is known but typically overlooked (see, for example, discussion in \citet{d19_sn2p} or \citet{goldberg_sn2p_19} for type II-P SNe or \citealt{D15_WR} for type Ic SNe). As we shall see, some of the degeneracy in interacting SN light curves can be lifted by using the additional information from spectra.

In the grid of \heracles\ simulations corresponding to Cases (1) and (2)  (see Section~\ref{sect_init}), we have identified one model in Case (1) and one in Case (2) that yield essentially the same bolometric light curve (top-right panel of Fig.~\ref{fig_couple}) despite the very different CSM structures adopted for each configuration (top-left panel of Fig.~\ref{fig_couple}). The case-(1) model corresponds to a CSM with an atmosphere with $H_\rho$ of 0.01\,$R_\star$ transitioning  into a wind with a mass-loss rate of $10^{-2}$\,\msunyr, with a total CSM mass of 0.024\,\msun, a total optical depth of 407 and a photospheric radius of  $3.12 \times 10^{14}$\,cm (equivalent to 4482\,\rsun; here, we assumed an ionized CSM, with $\kappa=$\,0.34\,cm$^2$\,g$^{-1}$) . The case-(2) model corresponds to a CSM with an atmosphere with $H_\rho$ of 0.5\,$R_\star$ transitioning  into a wind with a mass-loss rate of $10^{-5}$\,\msunyr, with a total CSM mass of 0.28\,\msun\ (hence ten times greater than in the case-(1) model), a total optical depth of 6770 and a photospheric radius of  $7.25 \times 10^{13}$\,cm (equivalent to 1042\,\rsun). This radius is very close to the outer edge of the high-density region and the wind itself is essentially optically thin. The CSM in the second case is therefore ten times more massive and about twenty times more optically thick but it yields a similar light curve as obtained for the first model. A comparison of \heracles\ light curves for the full set of simulations for case (1) and case (2) is presented in the appendix, in Fig.~\ref{fig_lbol_hd5em1_hd1em2}.

The reason for this similarity in light-curve properties in spite of the distinct CSM properties has to do with the different behavior of the shock in each configuration. To help visualize the differences, the bottom three rows of Fig.~\ref{fig_couple}  illustrate the evolution of the luminosity, density, and velocity in the first 4.5\,d after the start of the \heracles\ simulation in each case (one per column).

In the first case of a 0.024\,\msun\ CSM mass (left column, bottom three rows), a radiative precursor starts shortly after the shock crosses $R_\star$ and goes down the steep density decline of the atmosphere. This precursor is seen as a ``radiative tongue'' eating through the CSM ahead, causing a prompt ionization of this cocoon of material. In the figure, this is seen as the rising luminosity ahead of the shock in the first three epochs shown (left column, second panel from top) and it results from the leakage of photons stored behind the shock. This leakage starts at the shock when the radiative diffusion time becomes comparable to the shock crossing time through that overlying CSM (see, e.g., \citealt{falk_sbo_78}, \citealt{klein_chevalier_78}, \citealt{ensman_sbo_92}, \citealt{gezari_sbo_08}). As this ionization front raises the temperature and ionization of the initially cold CSM (i.e., 2000\,K), the material becomes optically thick to the incoming radiation. The optical depth in the Lyman continuum is large at all times but with the rising ionization caused by the precursor, electron scattering alone leads to optically-thick conditions for all photons whatever their wavelength. One day after the start of the simulation, the CSM is optically thick out to a radius of $3.12 \times 10^{14}$\,cm, as estimated above. At that time, the shock is still embedded within the CSM and at a radius of about $10^{14}$\,cm. As long as the shock is embedded within the slow-moving CSM, the photosphere does not move in radius and some unshocked CSM is present to reprocess the incoming radiation from the shock. It is during this phase that narrow symmetric lines broadened by electron scattering are present (the spectral evolution is analogous to that obtained for model r1w6 in \citealt{d17_13fs}; see also next section). This is the critical phase and configuration that lies behind the type IIn classification.\footnote{Some low-energy events may be classified as such just because they have intrinsically narrow lines so this classification requires caution.} The shock reaches the photosphere at 3.50--3.75\,d after the start of the simulation, at which time the photosphere (the dot in the bottom-left panel of Fig.~\ref{fig_couple}) suddenly speeds up to about 7000\,\kms\ as it is swept up by the dense shell (this shell is relatively cold and generally named a cold dense shell; CDS). This dense shell forms from the accumulation of swept-up CSM and decelerated ejecta. As the shock plows through the CSM (mass-loss rate of $10^{-2}$\,\msunyr), interaction shock power is produced and radiated. It contributes to producing the higher-than-standard SN luminosity and this leakage of radiative energy saps the total energy budget, which otherwise would have served to raise the kinetic energy of the ejecta. Indeed, at shock breakout, only about 50\% of the total energy below $R_\star$ is kinetic, while the other half is radiative. Hence, the CSM is the agent that drains the ejecta energy (first with the radiative precursor, then with the ejecta interaction with CSM), reducing in the process the asymptotic kinetic energy (see next section). Figure~\ref{fig_couple} shows that the maximum ejecta velocity in this model is about 9000\,\kms\ at 1\,d (in the absence of the radiative leakage at breakout this velocity would have been 11000\,\kms), and then decreases until the last snapshot here at 4.50\,d to about 8500\,\kms\ as the ejecta continue to plow through the relatively dense outer CSM.

In the second case of a 0.28\,\msun\ CSM mass (right column, bottom three rows), the behavior of the shock is very different. As the shock crosses $R_\star$ and enters the high-density CSM region, no precursor forms. The density is so high that the photon mean free path $1/\kappa \rho$ remains small compared to the local radius $R$ (i.e., $\lesssim$\,1\,\rsun\ compared to 500-1000\,\rsun). Photons remain trapped behind the shock and the radiative precursor is delayed until the shock arrives at the outer edge of the dense CSM, at about 1000\,\rsun\ in this model. The subsequent evolution is then as in the previous case except that now the wind mass-loss rate is small and the outer CSM is optically thin. There is thus no optically-thick CSM to reprocess the radiation incoming from the shock and no narrow line typical of type IIn SNe can arise, except perhaps for an hour or so. From the point of view of type IIn signatures, this model is analogous to one in which there was no CSM (see, for example, model r1w1 in \citealt{d17_13fs}). However, from the point of view of the bolometric light curve, this model has a significant boost at early times because the energy left over by the shock in the dense CSM can escape on a week time scale -- this diffusion time is essentially ten times shorter than for the layers below $R_\star$ because the density is ten times smaller. This phenomenon is analogous to that pertaining to type IIb SNe from progenitors with extended envelopes (e.g., SN\,1993J; see, for example, \citealt{woosley_94_93j}; \citealt{d18_ext_ccsn}). The absence of shock breakout deep into an extended CSM implies that there is no long-lived type IIn signature, but the lack of a precursor also implies that there is no radiative leakage during the shock crossing of the CSM. The shock propagates adiabatically all the way to the outer edge of the CSM. The lack of radiative leakage thus explains the higher maximum velocity reached in this model (about 11000\,\kms, bottom-right panel of Fig.~\ref{fig_couple}), of a similar magnitude as obtained in the case of no CSM.

To conclude, the two different models discussed here correspond to completely different interaction configurations because the shock behavior is strongly affected by the adopted CSM density. At higher CSM densities, the shock propagates through the CSM without radiative losses, and thus adiabatically, as when it was crossing the envelope of the progenitor star. This impacts the radiative precursor, the breakout signal, the luminosity boost, the appearance, existence, and persistence of narrow spectral lines formed in the unshocked CSM. These differences hold despite the very similar bolometric light curves. In addition, the model with the lower CSM density yields the same bolometric boost despite the factor of 10 lower CSM mass. This suggests that CSM masses inferred from light-curve modeling (see, for example, \citealt{morozova_sn2p_18,haynie_sbo_21,moriya_sn2p_23,subrayan_sn2p_23}) are not uniquely defined and potentially strongly overestimated. A critical observable that must be used for constraining more accurately the CSM density and extent (and thus contribute to determining the CSM mass) is the spectral evolution and in particular the survival time of type IIn signatures (see, for example, \citet{leonard_98S_00} for SN\,1998S,\citet{yaron_13fs_17} for SN\,2013fs,  or \cite{terreran_20pni_22} for SN\,2020pni).

\begin{figure}
\centering
\includegraphics[width=0.9\hsize]{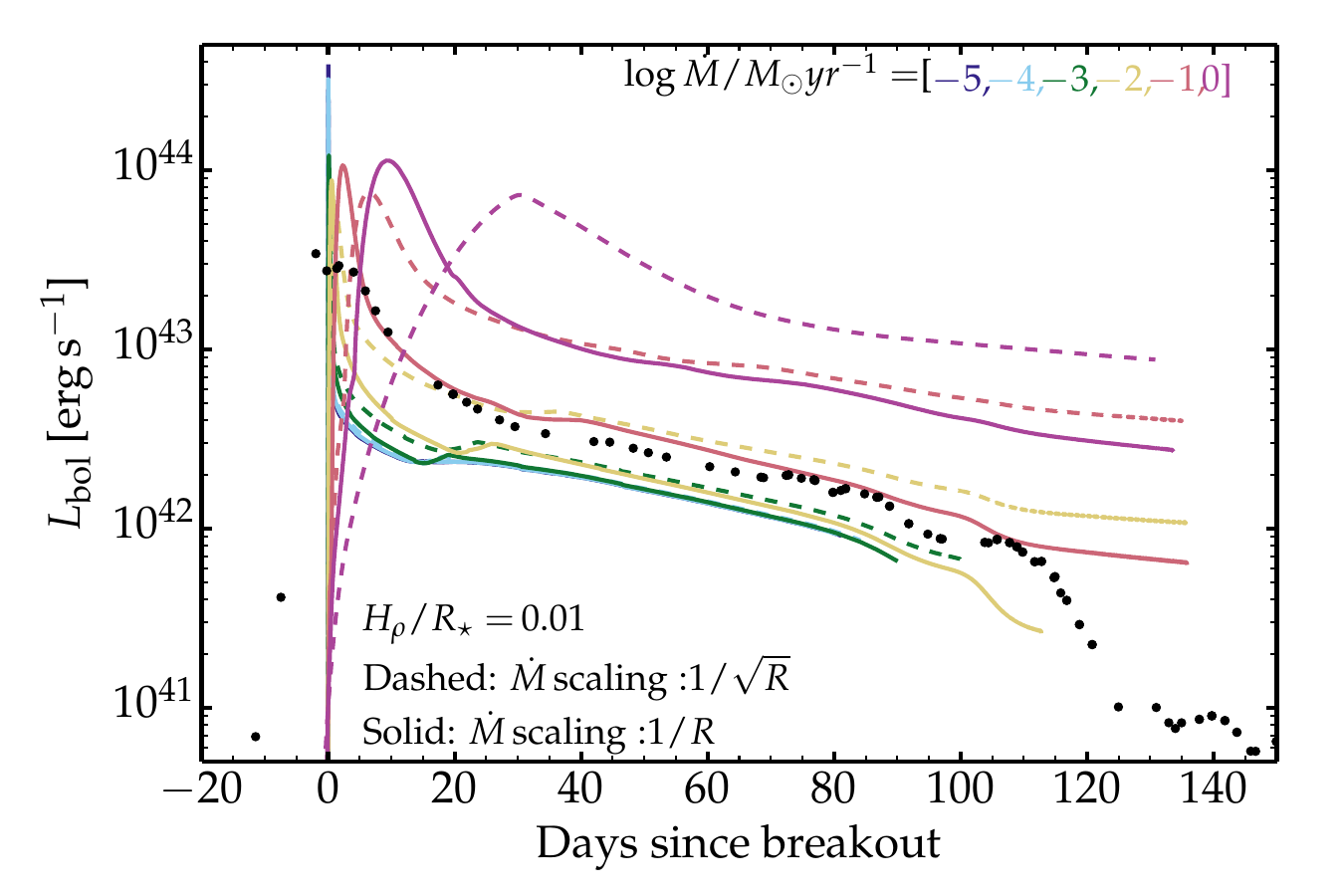}
\includegraphics[width=0.9\hsize]{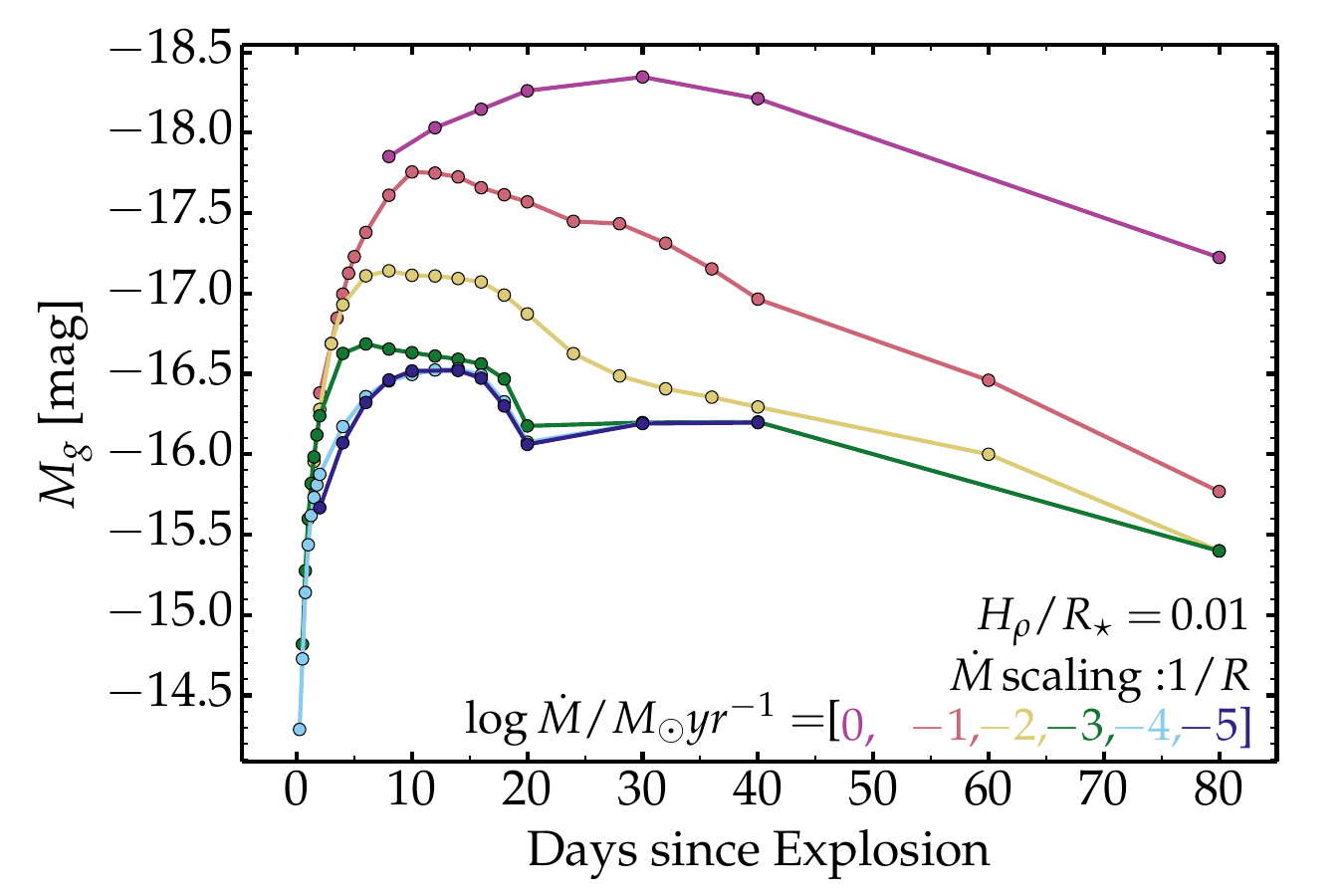}
\includegraphics[width=0.9\hsize]{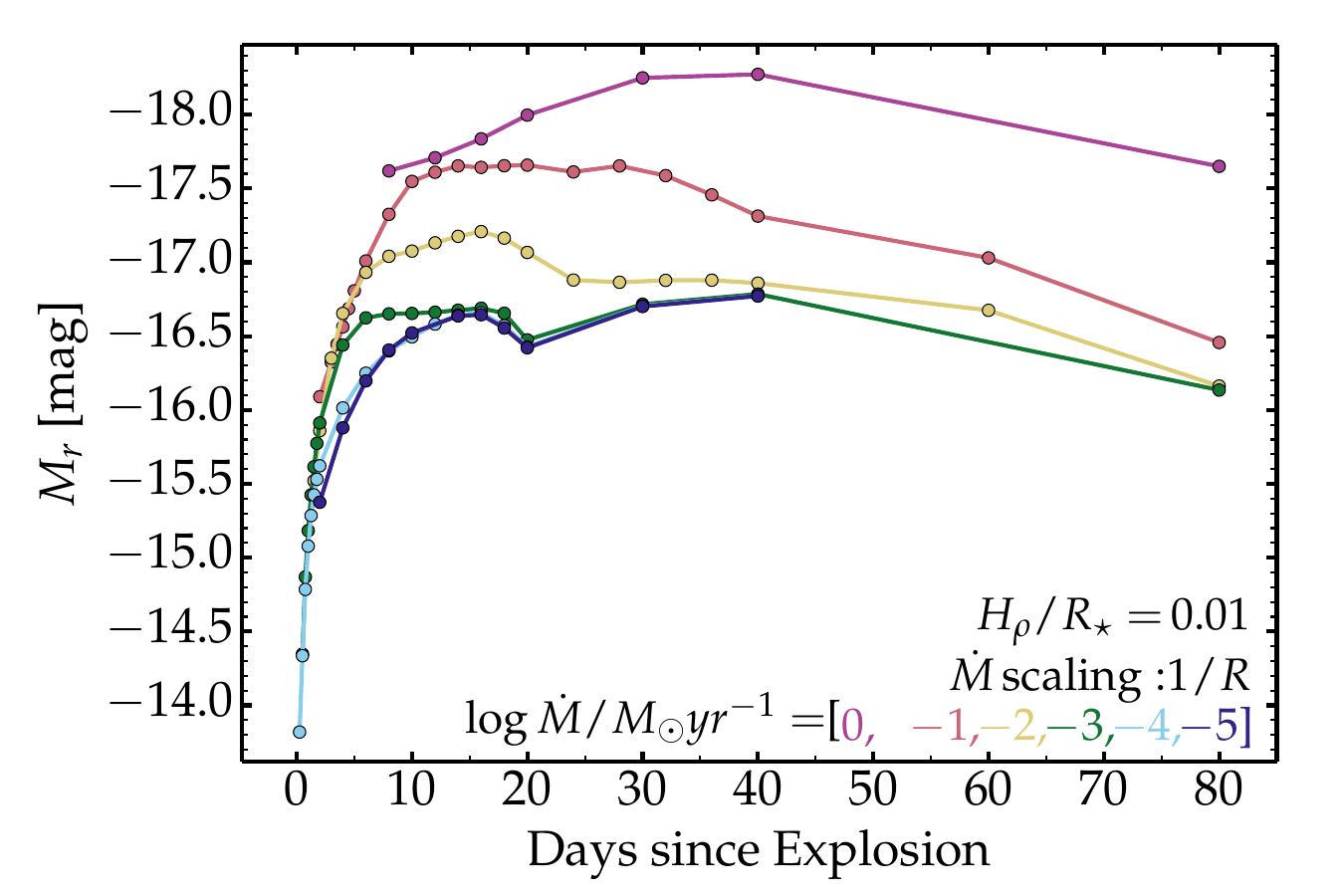}
\caption{Photometric properties of simulations corresponding to case-(1) CSM configurations. Top panel: \heracles\ bolometric light curves for wind mass-loss rates covering from 10$^{-5}$ up to 1\,\msunyr\ with an additional density scaling of $1/R$ (solid) and  $1/\sqrt{R}$ (dashed). Filled dots correspond to the inferred values for SN\,2020tlf \citep{wynn_20tlf_22}. Simulations using a lower wind density are truncated at late times because of a numerical problem when the shock reaches the outer grid boundary. Middle and bottom panels: $g$ (middle) and $r$-band (bottom) light curves computed with \cmfgen\ and based on the \heracles\ simulations for interaction models with an atmospheric scale height of 0.01\,$R_\star$, wind mass-loss rates between 10$^{-5}$ and 1\,\msunyr, and an additional scaling of the wind density by a factor of $1/R$.
\label{fig_phot}
}
\end{figure}

\begin{figure*}
\centering
\includegraphics[width=0.35\hsize]{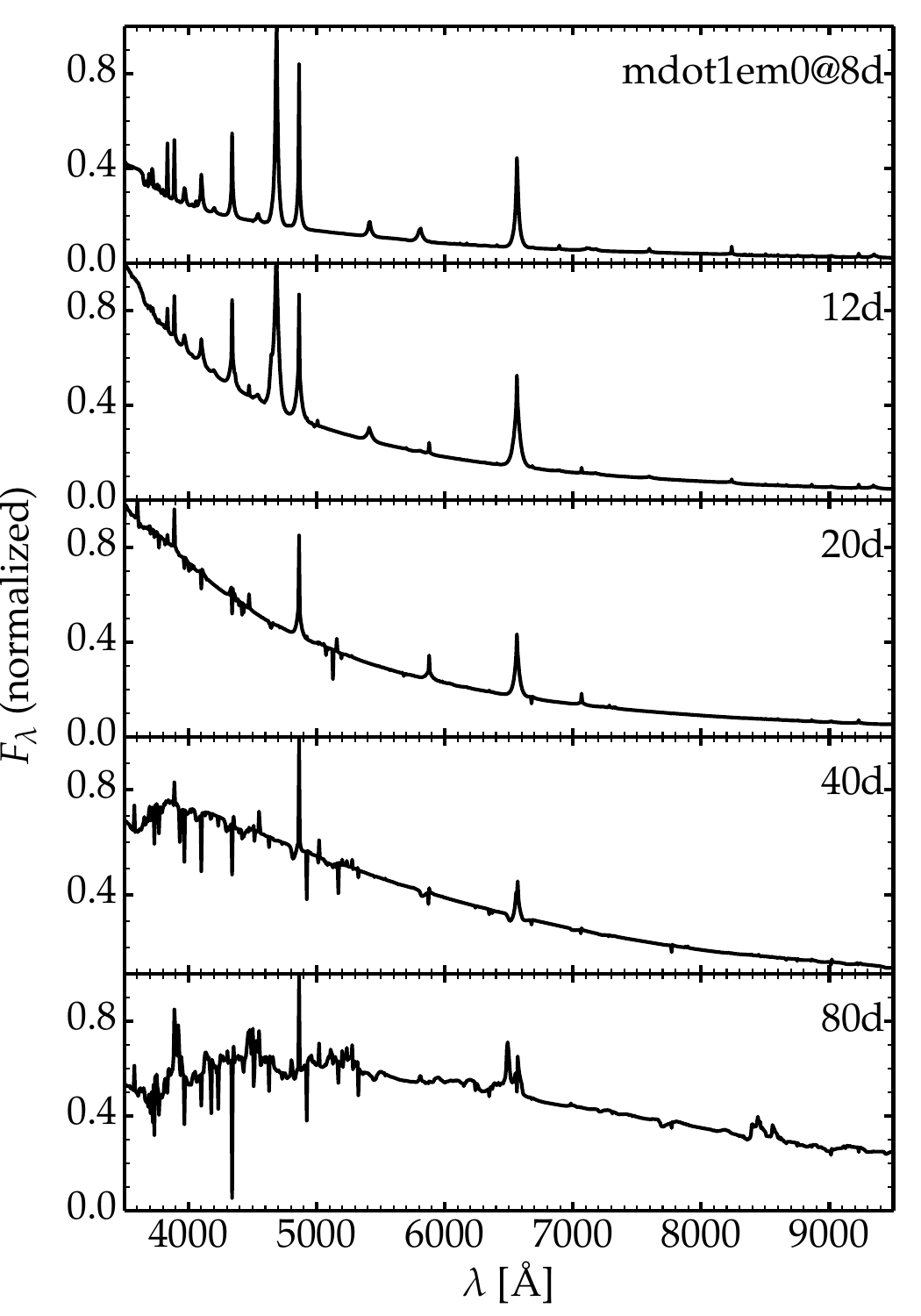}
\includegraphics[width=0.35\hsize]{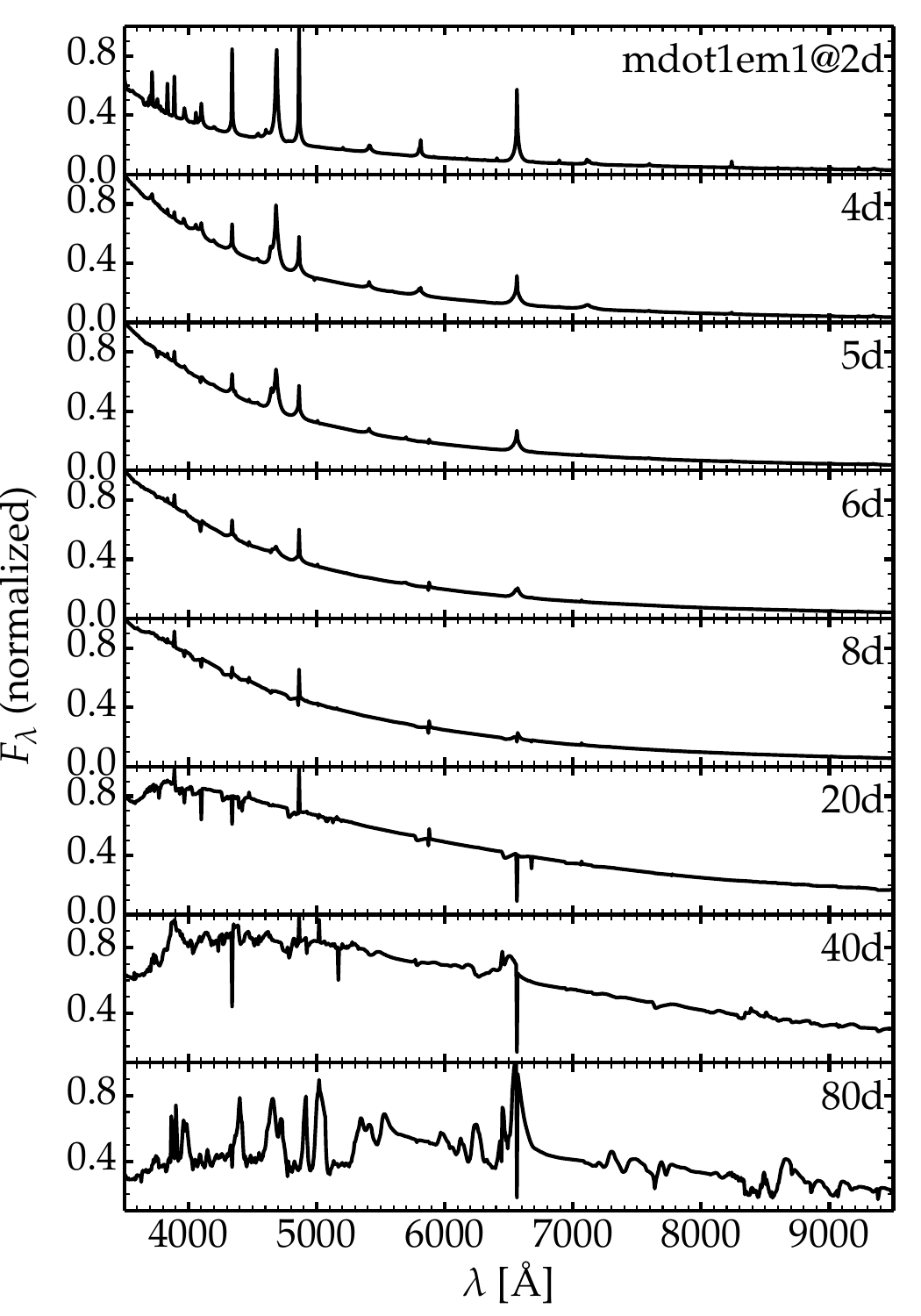}
\includegraphics[width=0.35\hsize]{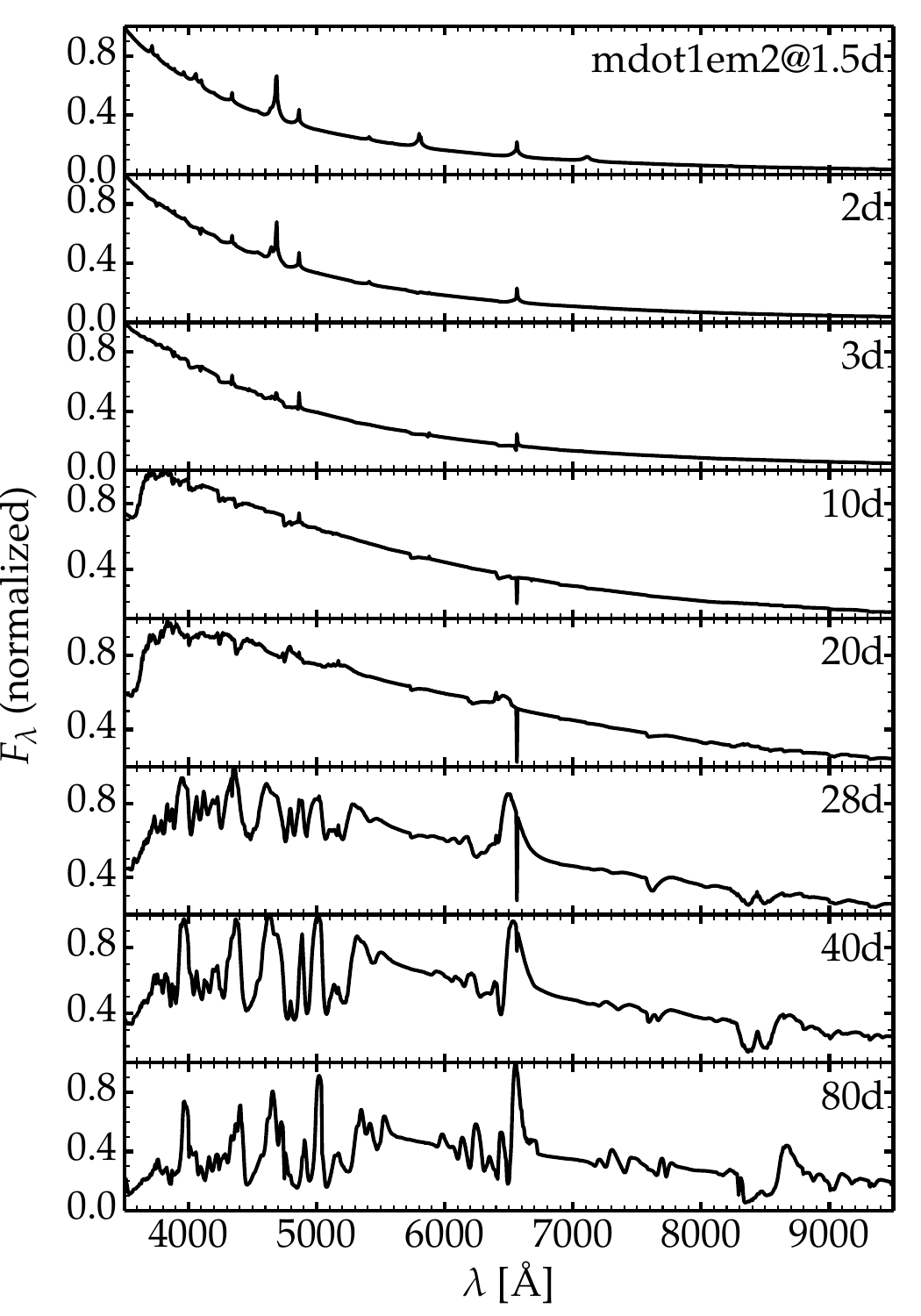}
\includegraphics[width=0.35\hsize]{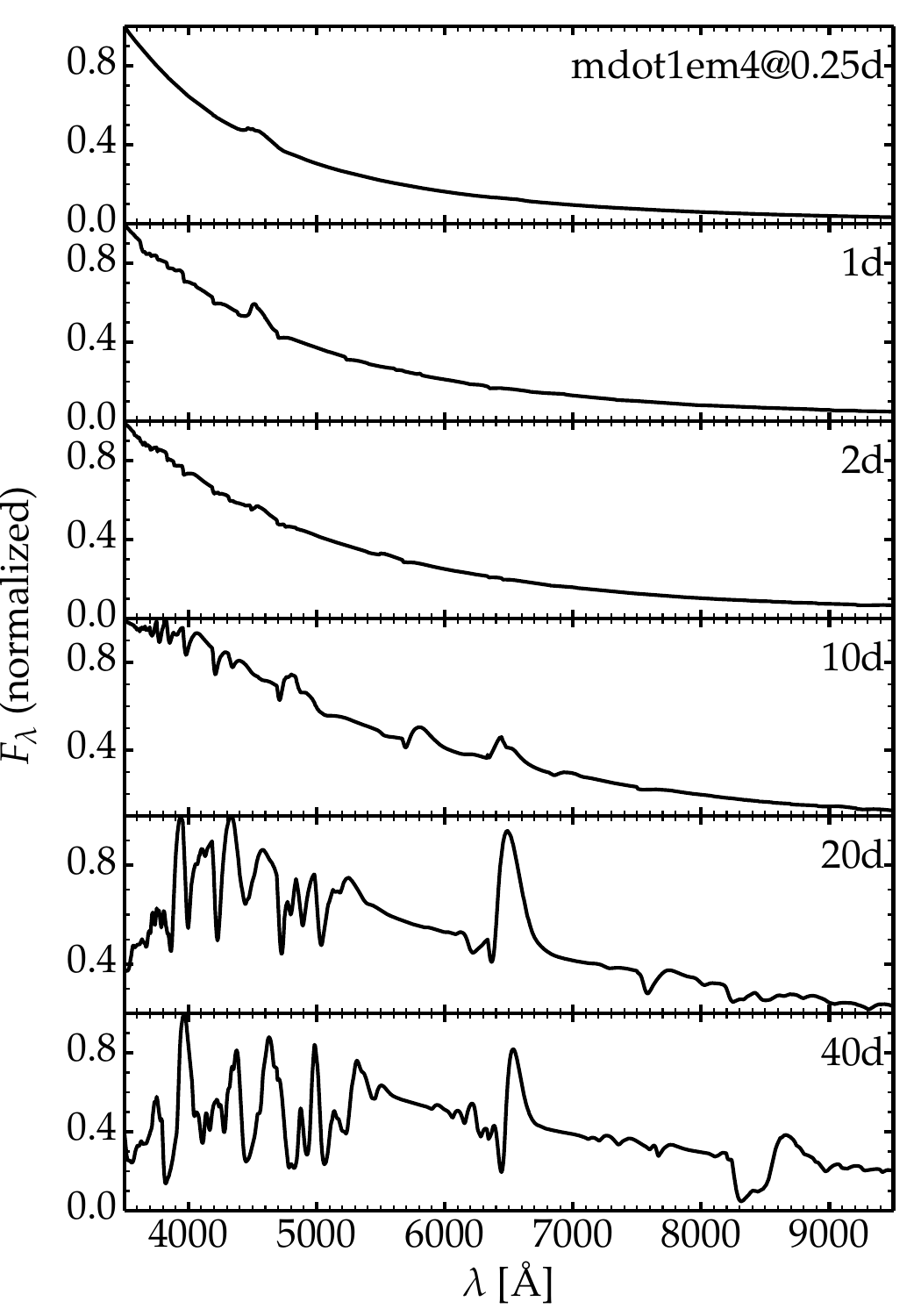}
\caption{Multiepoch model spectra for various CSM configurations. We show the spectral evolution computed with \cmfgen\ and based on the \heracles\ simulations for the case-(1) CSM models (atmospheric scale height of 0.01\,$R_\star$) with wind mass-loss rate of 1.0 (top left; model mdot1em0), 0.1 (top right; model mdot1em1), 0.01 (bottom left; model mdot1em2), and 0.0001\,\msunyr\ (bottom right; model mdot1em4). An additional scaling of the wind density by a factor of $1/R$ is employed.
\label{fig_spec_mdot}
}
\end{figure*}

\section{The diversity of type II-P SNe with early-time type IIn signatures}
\label{sect_grid}

We now discuss a number of correlations between gas and radiation properties associated with a range of CSM mass, density, or extent. As we discussed above, the high CSM density, analogous to stellar interiors, corresponding to case (2) above does not lead to any IIn spectral signatures, which are the essence of events like SN\,2013fs. Such CSM configurations are therefore no longer considered. Instead, we consider the case-(1) CSM configurations corresponding to a CSM of lower mass or density (the density profiles considered are shown in the top panel of Fig.~\ref{fig_init}). Using our grid of \heracles\ and \cmfgen\ simulations, we discuss the correlations between gas, photometric, and spectroscopic properties, and in particular in relation to the presence of IIn signatures in optical spectra. Since \citet{d17_13fs} already performed a detailed study of that topic, we first briefly discuss the results from the \heracles\ and \cmfgen\ simulations before presenting in more detail the important correlations that result.

\subsection{Photometric and spectroscopic properties}
\label{sect_res}

Figure~\ref{fig_phot} presents some photometric properties for the case-(1) CSM configurations. The bolometric luminosity light curves from \heracles\ (top panel) cover from standard type II SN luminosities for low mass-loss rate up to high luminous events (still short of superluminous SNe IIn) for the highest mass-loss rates, as expected. For a higher wind density, the luminosity peak is broadened, both as a consequence of the diffusion-broadening of the breakout burst and the continued supply of interaction power as the shock crosses the CSM. This also occurs for curves corresponding to the same nominal mass-loss rate but different additional radial scaling of the density (i.e., $1/R$ instead of $1/\sqrt{R}$). These models straddle the inferred luminosity from SN\,2020tlf, shown as dots in Fig.~\ref{fig_phot}. Any luminosity offset at $>$\,100\,d reflects primarily the overestimated contribution from the persisting interaction with CSM and so could be resolved by adjusting the outer properties of the CSM -- this issue is irrelevant for the present discussion.

The bottom panels in Fig.~\ref{fig_phot} illustrate the $g$ and $r$-band light curves for the case-(1) models with a $1/R$ additional density scaling. Although the curves show some breaks in cases,\footnote{These breaks are associated, for example, with the sudden change caused by the recession of the photosphere through the dense shell -- see Fig.~\ref{fig_couple}. This artifact results largely from the huge density variation across the dense shell inherent to the assumption of spherical symmetry.} the influence of CSM is to reduce the rise time in optical bands as long as the wind density is not too large. However, as the wind density is increased to larger values, the rise time eventually increases because of the increase of the radiative diffusion time associated with the CSM. This trend agrees with previous modeling work (see, for example, \citealt{moriya_rsg_csm_11}) and the observations reported by \citet{bruch_csm_21}. The actual peak magnitudes depend on the adopted explosion energy or progenitor radius. A higher CSM density (for example near $R_\star$) may also boost the luminosity (see Section~\ref{sect_degen}).

\begin{figure*}
\centering
\includegraphics[width=0.45\hsize]{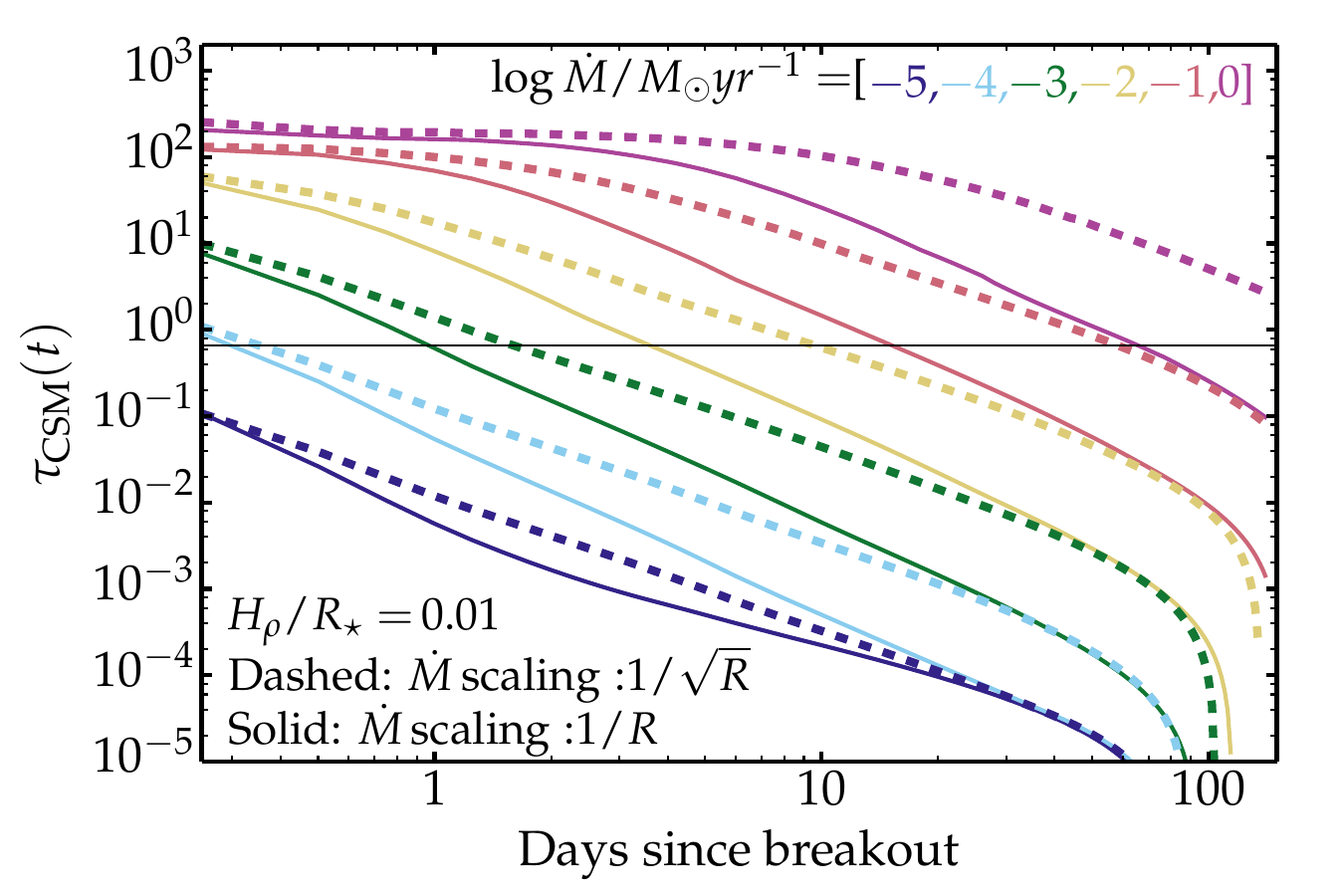}
\includegraphics[width=0.45\hsize]{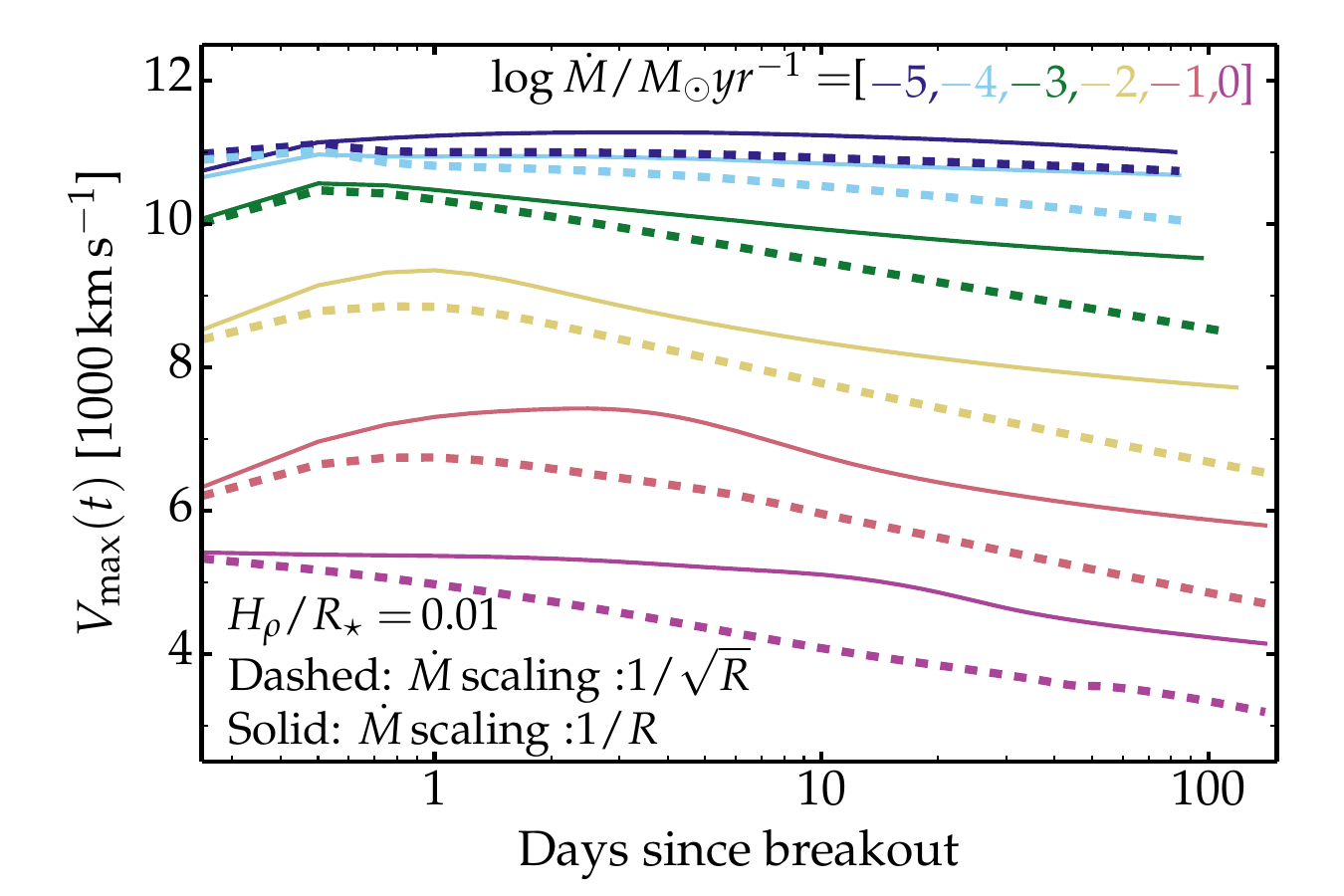}
\includegraphics[width=0.45\hsize]{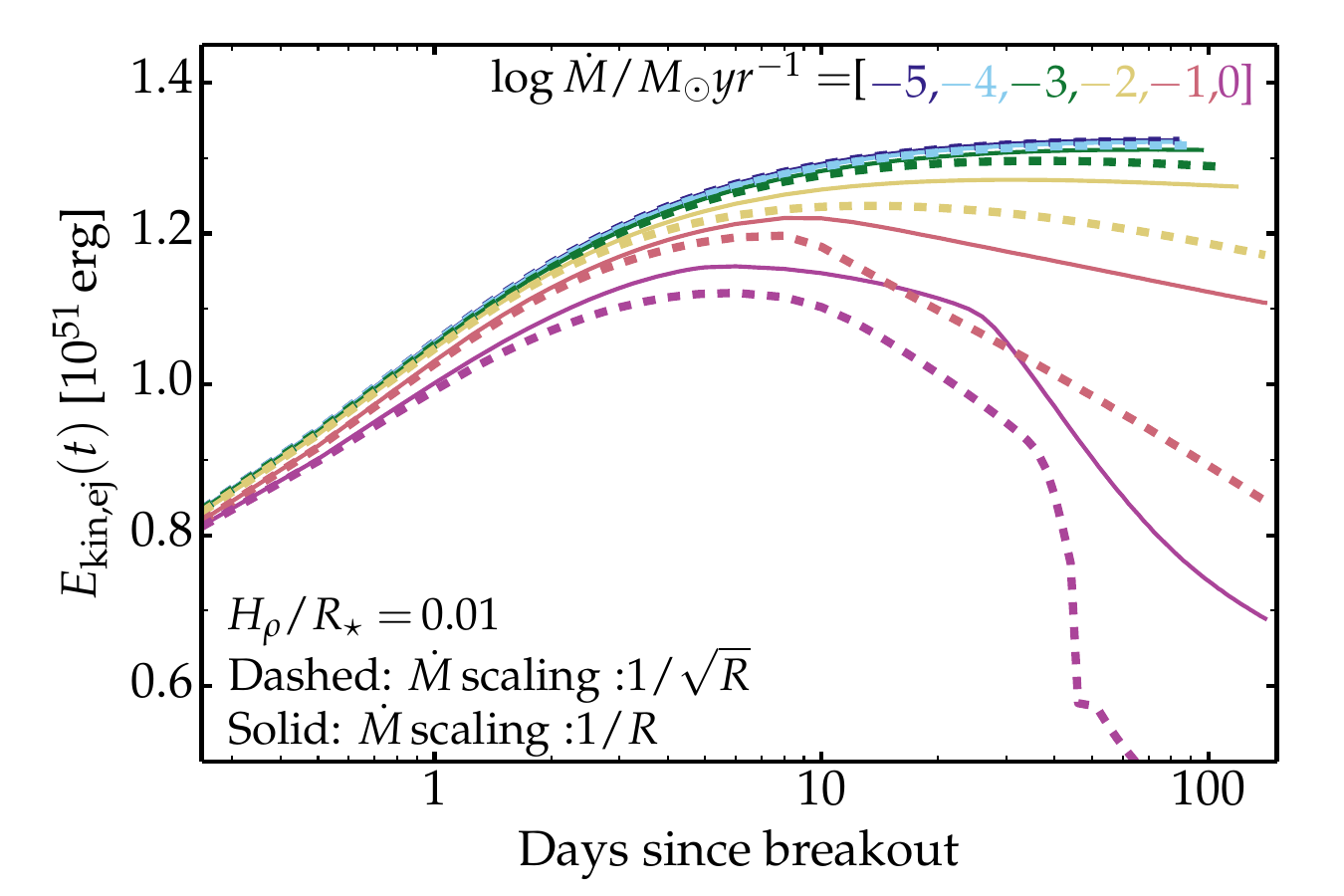}
\includegraphics[width=0.45\hsize]{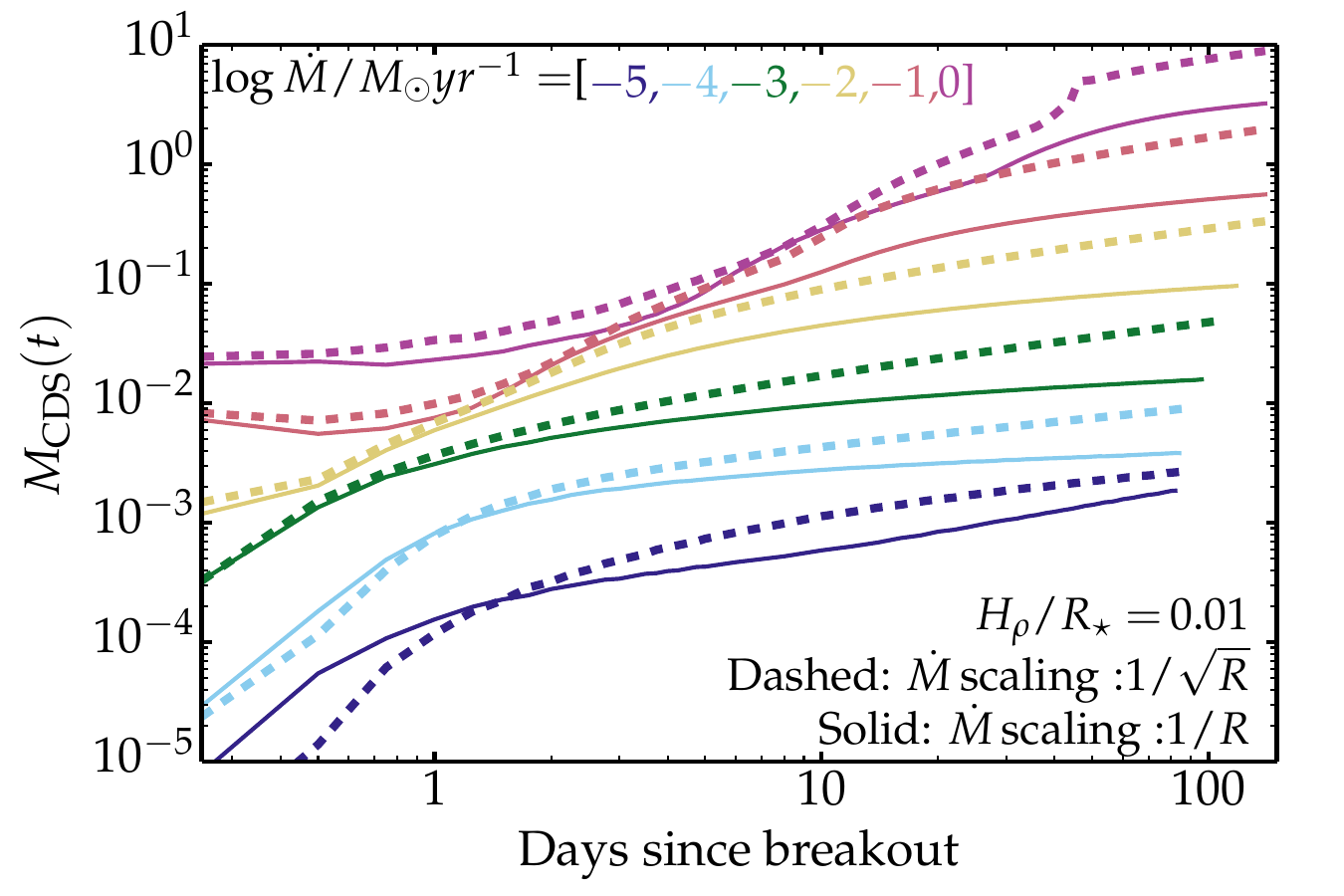}
\includegraphics[width=0.45\hsize]{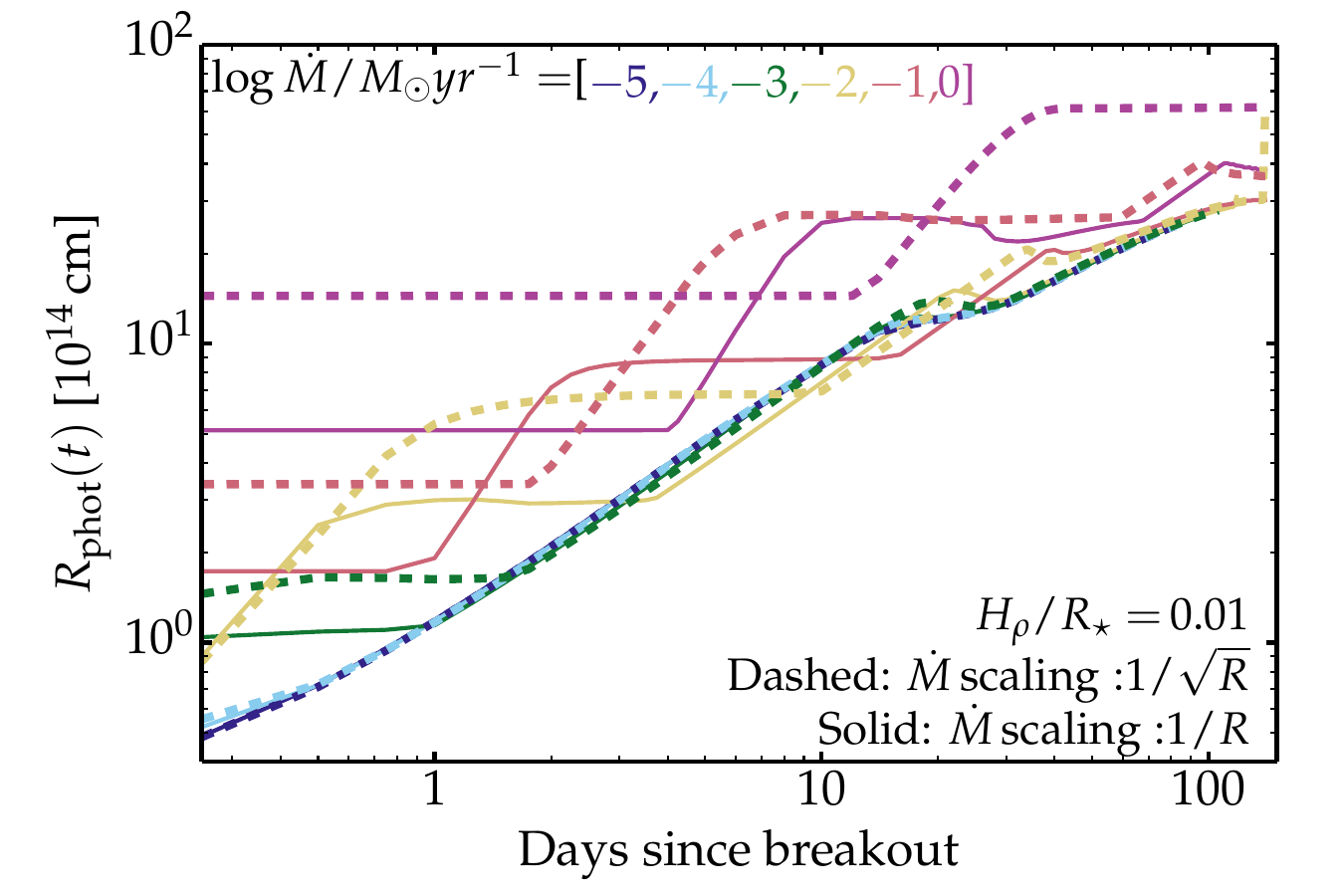}
\includegraphics[width=0.45\hsize]{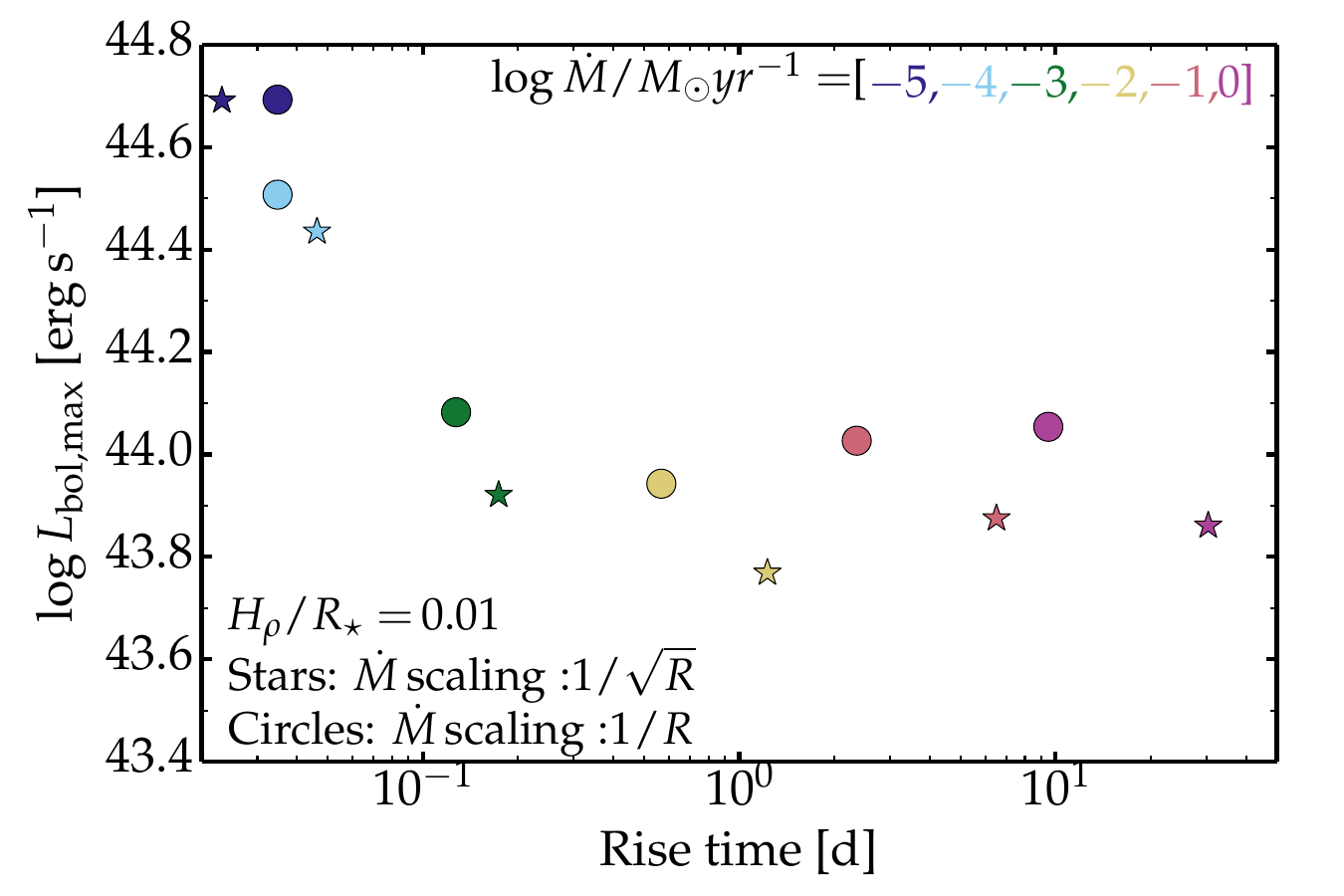}
\caption{Properties of simulations for progenitors with wind-like CSM covering a range of mass-loss rates and different additional radial scaling. We show the time evolution of the electron-scattering optical depth of the unshocked CSM (top left), the time evolution of the maximum ejecta velocity (top right), the time evolution of the ejecta kinetic energy (middle left), the time evolution of the mass contained in the CDS (middle right), the time evolution of the photospheric radius (bottom left), and the correlation between the time and luminosity at bolometric maximum (bottom right).
\label{fig_glob_prop}
}
\end{figure*}

Figure~\ref{fig_spec_mdot} illustrates the spectral evolution for Case-(1) CSM models with a range of wind mass-loss rates. We only show four representative cases to avoid redundancy. For models with a high wind density, the spectral evolution is qualitatively similar to that obtained earlier for models \citep{D16_2n} of the well-observed SN\,1998S \citep{leonard_98S_00} in the sense that we identify three distinct phases. During the first phase, the models exhibit a blue continuum with emission lines characterized by a narrow core and symmetric wings broadened by electron scattering. Here, the model predicts a combination of lines from H and He primarily, with for example the Balmer series especially with strong H$\alpha$ and H$\beta$ as well as lines of He\one\ at 5875, 6678, and 7065\,\AA, and of He\two\ at 4686 and 5411\,\AA.\footnote{Together with He\two\,4686\,\AA,  observations generally exhibit another emission line due to N\three\,4634\,\AA, and more rarely, at higher temperature and ionization, N\five\,4604--4620\,\AA. These lines are, however, difficult to model \citep{D16_2n,d17_13fs}. Here, N\three\,4634\,\AA\ is predicted in the model with 10$^{-1}$\,\msunyr\ at 4 and 5\,d (top-right panel of Fig.~\ref{fig_spec_mdot}), and in the model with 10$^{-2}$\,\msunyr\ at $\sim$2\,d (bottom-left panel of Fig.~\ref{fig_spec_mdot}).} The spectrum forms in the unshocked CSM and the underlying fast moving material is entirely obscured. This first phase lasts for about 30\,d in the model with the highest wind mass-loss rate, about 7\,d in the model with 10$^{-1}$\,\msunyr, and about 2\,d in the model with 10$^{-2}$\,\msunyr. Models with a low wind density (nominal mass-loss rate of 10$^{-4}$\,\msunyr\ or lower, which also include a $1/R$ density scaling), the spectral evolution is essentially identical to the case of no CSM at all (bottom-right panel in Fig.~\ref{fig_spec_mdot}). The type IIn signatures are not even present in the first spectrum at 0.25\,d after the start of the simulation. This latter model does exhibit a blue continuum but only one line is present at 0.25\,d, which is associated with blueshifted, Doppler-broadened emission from He\two \,4686\,\AA. In this case the spectrum forms in the fast moving ejecta and the wind is too tenuous to cause any significant reprocessing of the incoming SN radiation. The observation of blueshifted broad He\two \,4686\,\AA\ is a natural consequence of the high temperature and high velocity of the type II SN photosphere and does not require prior or ongoing interaction with CSM. This Doppler-broadened He\two\ line is observed in many young SNe II (e.g., SN\,2018lab, \citealt{pearson_18lab_23}) and should not be interpreted as CSM interaction (e.g., \citealt{bruch_csm_21}).

In the higher wind density models, the second phase shows an essentially blue featureless spectrum without any narrow, type IIn-like emission lines but instead   absorptions that are blueshifted by several 1000\,\kms. The spectrum at that phase forms in the dense shell at the junction between ejecta and CSM. This dense shell, which contains the swept-up CSM and the decelerated ejecta, travels fast (essentially at the maximum velocity of the ejecta) and is at the origin of the Doppler shift of the absorption in spectral lines -- there is no optically-thick, slow (i.e., material that was either not shocked or not accelerated by the SN radiation)  CSM to produce electron-scattering broadened lines.

In the third phase, the spectrum forms in the unshocked ejecta, with a modest contribution from the outer dense shell -- this phase starts essentially at shock breakout for the low wind-density  cases (nominal mass-loss rate of 10$^{-4}$\,\msunyr\ or lower). Here, the combined modeling with \heracles\ and \cmfgen\ shows some limitations since the NLTE effects associated with the shock are not properly modeled (this is largely caused by the fixed temperature in the \cmfgen\ calculation). During that phase, a better approach is to treat the interaction shock power directly into \cmfgen, as in \citet{dessart_csm_22}. With that approach, interaction power absorbed in the dense shell leads to the formation of broad boxy emission profiles.

\subsection{Correlations}
\label{sect_corr}

The presence of CSM at the surface of exploding RSG stars impacts the radiative properties of type II-P SNe both for the photometry (fainter and longer-lived shock breakout burst, early-time luminosity boost, shorter optical rise time) and for the spectroscopy (presence of narrow-core electron-scattering broadened lines formed in unshocked CSM). In addition, these properties are directly related to modifications of the ejecta structure and dynamics, resulting from the radiative losses (prolonged radiative precursor), the interaction with CSM, and the extraction of kinetic energy. Below, we review a number of such quantities and connect them to observables. Such correlations between distinct observables offer a means to lift some of the degeneracies affecting light curves or spectra.

Figure~\ref{fig_glob_prop} illustrates the evolution since shock breakout (to be unambiguous, we mean the time at which the shock crosses $R_\star$; see Sect.~\ref{sect_init} and Fig.~\ref{fig_init}) for the unshocked CSM electron-scattering optical depth $\tau_{\rm CSM}(t)$ (top left), the maximum velocity on the grid $V_{\rm max}(t)$ (which corresponds to the ejecta material immediately interior to the ejecta-CSM interface; top right), the ejecta kinetic energy $E_{\rm kin, ej}(t)$ (middle left), the accumulated mass in the dense shell $M_{\rm CDS}(t)$ (middle right), and the radius of the electron-scattering photosphere $R_{\rm phot}(t)$ (bottom left). The bottom-right panel shows the correlation between the peak luminosity and the bolometric rise time. In all panels, we show the results for the case-(1) CSM models with both a $1/R$ and a $1/\sqrt{R}$ scaling to the CSM density. The results are essentially the same except that a flatter declining CSM density acts in a similar way to a higher wind mass-loss rate. Hence, we only discuss one case (i.e., simulations with a $1/R$ additional density scaling).

As expected, the presence and survival time of type IIn signatures (see spectral sequences in Fig.~\ref{fig_spec_mdot} and Sect.~\ref{sect_res}) directly correlates with $\tau_{\rm CSM}$  (top left panel of Fig.~\ref{fig_glob_prop}). Our models with a nominal wind mass-loss rate greater than 10$^{-3}$\,\msunyr\ exhibit type IIn signatures for at least $\sim$\,1\,d, and these signatures can persist from weeks to months, as long as there is optically-thick unshocked CSM (i.e., $\tau_{\rm CSM}>2/3$). In the model with the highest mass-loss rate, $\tau_{\rm CSM}=2/3$ occurs at $\sim$\,60\,d. Such a long duration is possible only because the CSM density is large out to large radii (but the CSM mass is modest so the density is not too large) so that it takes many weeks for the shock to overtake that buffer of mass.

An obvious impact of CSM is the reduction of $V_{\rm max}(t)$ (top right panel of Fig.~\ref{fig_glob_prop}). When CSM is placed at the surface of the exploding star, it first saps the radiative energy stored behind the shock. This leakage reduces the amount of energy available to accelerate the ejecta. Subsequently, as the shock plows through CSM, more radiation leaks out, and a direct deceleration of the ejecta occurs through interaction with the CSM. For the cases with a low wind density, this reduction in $V_{\rm max}$ is small and $V_{\rm max}$ reaches a value of about 11000\,\kms. For increasing CSM density and mass, the reduction increases, leading to very small values in the strongest interacting cases ($V_{\rm max}$ converges to about 4000\,\kms\ in the highest mass-loss rate model). Another aspect is that the $V_{\rm max}(t)$ is indeed a function of time, that is the ejecta-CSM interaction continuously leads to ejecta deceleration so that one can never strictly assume a homologous ejecta expanding in a vacuum like in normal type II-P SNe.

The impact of CSM on $V_{\rm max}(t)$ has a counterpart in the profile of $E_{\rm kin,ej}(t)$ (middle-left panel of Fig.~\ref{fig_glob_prop}). For low wind mass-loss rates, the ejecta kinetic energy is $1.32 \times 10^{51}$\,erg, which is within 1.5\% of the value predicted for model m15mlt3 without CSM in \citet{d13_sn2p}. This ``asymptotic" kinetic energy is reached after about ten days, even in the absence of CSM or for a weak CSM. However, with denser winds, we see that several 10$^{50}$\,erg can go ``missing" in just a few days as they have been radiated away, thereby boosting the luminosity of the SN above standard, type II-P SN values. CSM can therefore have a very detrimental impact on the SN energetics. It can indeed serve to boost the SN luminosity but this comes at the expense of the ejecta kinetic energy. A critical diagnostic is to check the maximum velocity of the ejecta, for example by examining the velocity at which the absorption troughs of lines like H$\alpha$ join with the continuum (this can only be done once the CSM is optically thin so that H$\alpha$ forms at least in part in the underlying unshocked ejecta). Any interaction should lead to a relative deficit of fast material, and this deficit should be more and more severe for a greater CSM mass.

Another consequence of ejecta-CSM interaction is the sweeping-up of CSM material and the accumulation of ejecta material into a dense shell at the interface between ejecta and CSM. The middle-right panel of Fig.~\ref{fig_glob_prop} shows the evolution of that accumulated mass $M_{\rm CDS}(t)$. At the end of the \heracles\ simulations, $M_{\rm CDS}(t)$ reaches 0.001\,\msun\ in cases of low wind density (i.e., 10$^{-5}$\,\msunyr), 0.1\,\msun\ for moderate wind density (i.e., 10$^{-2}$\,\msunyr), and exceeds 1\,\msun\ for the highest mass-loss rates. Typically, a CSM of a given mass decelerates a comparable mass of ejecta material, from which one may estimate the amount of kinetic energy extracted and the resulting maximum velocity (see discussion in \citealt{HD19}). Importantly, this outer dense shell may absorb a fraction of the shock power at late times and give rise, for example, to broad boxy emission profiles as observed in SN\,1993J \citep{matheson_93j_00b} or other historic type II SNe \citep{milisavljevic_late_12}. The presence of such broad emission features are a signature of ongoing interaction. However, it is expected even in cases of moderate wind mass-loss rates \citep{d13_late_sn2p}.

The evolution of the photospheric radius illustrates the complicated phenomena taking place in an ejecta-CSM interaction. This radius may not be easily identified in observed transients but in radiation hydrodynamics simulations, it is useful as it traces roughly the spectrum formation region and can thus explain the spectral properties. The rate of change of $R_{\rm phot}(t)$ does not, however, translate into a velocity at the photosphere. For example, an ionization front progressing through an unshocked, slow, constant-velocity CSM would yield an increasing $R_{\rm phot}(t)$ but a constant and small $V_{\rm phot}(t)$ (this is a matter of distinguishing the velocity of the photosphere and the velocity at the photosphere). The low wind density models are trivial to interpret because the photosphere expands with the ejecta at all times (it generally recedes in mass space though). The complications arise in the presence of a dense wind. The model with a $10^{-1}$\,\msunyr\ offers some interesting features. In this model, $R_{\rm phot}(t)$ is initially small and located within the cold CSM, above $R_\star$. As the radiative precursor crosses and ionizes the CSM, $R_{\rm phot}(t)$ steeply increases between $\sim$\,1\,d to $\sim$\,2d, essentially tracking the ionization front. It then stays constant at $\sim 9 \times 10^{14}$\,cm until $\sim$\,20\,d, when the CDS crosses that location. The photosphere is then located in the CDS and expands at about 6000\,\kms. The CDS then becomes optically thin and the photosphere recedes in mass space, down into layers that move slower -- the rate of change of $R_{\rm phot}(t)$ decreases (i.e., the curve flattens). Other simulations exhibit a similar behavior although the different phases appear shorter or longer depending on the shock trajectory and CSM properties.

The bottom-right panel of Fig.~\ref{fig_glob_prop} shows how the bolometric rise time correlates with the bolometric maximum. For low wind densities, there is a clear correlation in which higher wind density leads to an increase in rise time, which reflects the broadening of the breakout burst over a CSM diffusion time. For our models with a wind mass-loss rate  $\gtrsim10^{-2}$\,\msunyr, the bolometric maximum stays roughly constant but the rise time increases. In these cases, the time-integrated bolometric luminosity increases. In other words, a greater CSM density and mass tends to extract more kinetic energy and boost the radiative energy budget but because the CSM optical depth also increases, this energy is released on a longer time scale and the peak luminosity stays roughly unchanged.

\begin{figure}
\centering
\includegraphics[width=\hsize]{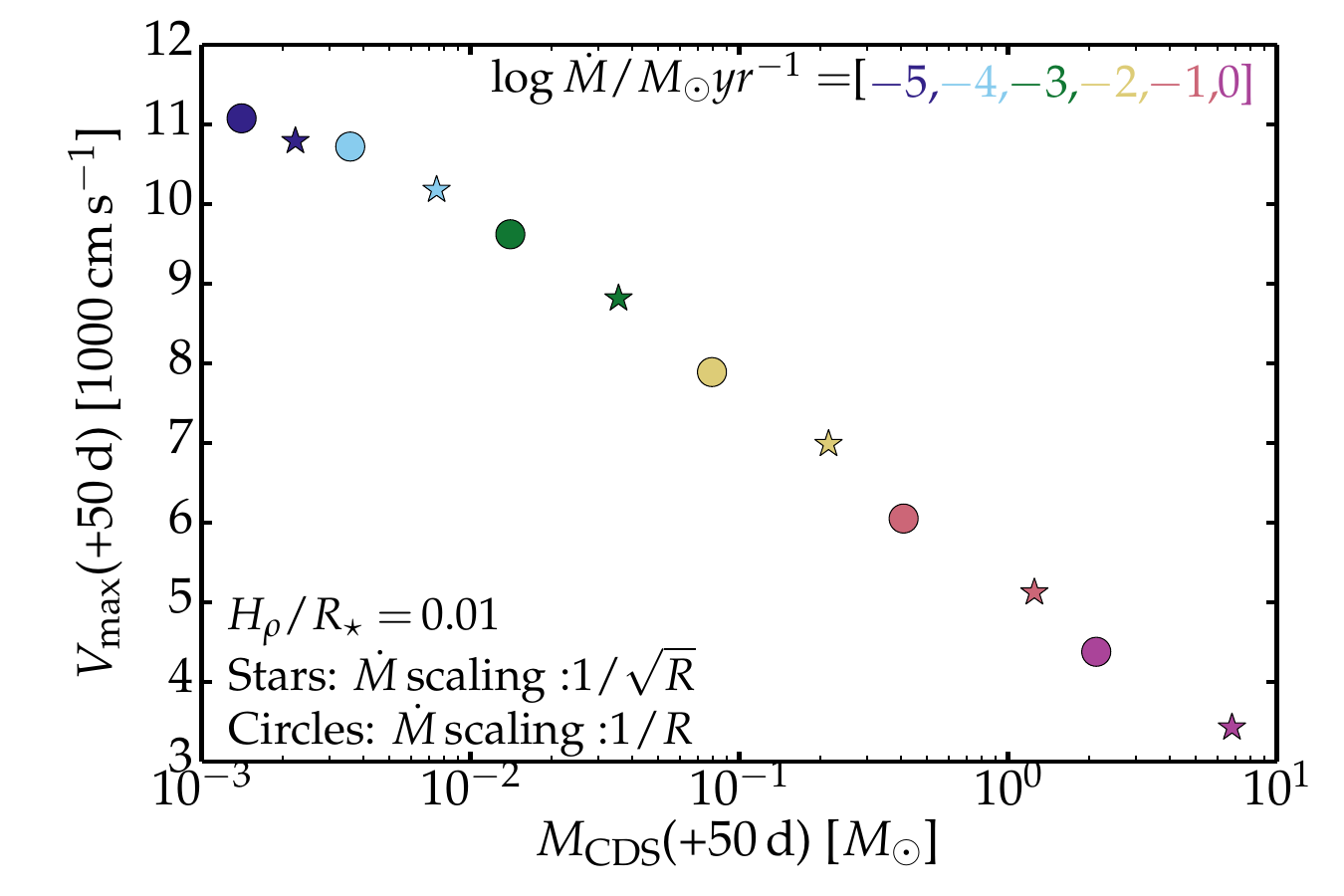}
\caption{Correspondence between the maximum ejecta velocity $V_{\rm max}$ (which is about 10\% greater than the CDS velocity) and the mass of the CDS $M_{\rm CDS}$, both at 50\,d past bolometric maximum. We show models with case-(1) CSM and the two different additional radial scalings for the density.
\label{fig_vmax_vs_mcds}
}
\end{figure}

Figure~\ref{fig_vmax_vs_mcds} illustrates one final correlation between $V_{\rm max}$ and $M_{\rm CDS}$ at 50\,d after breakout. The evolution of each quantity was  shown in Fig.~\ref{fig_glob_prop} but with respect to bolometric rise time. What is clearly apparent is that the two are strongly correlated and offer one means to constrain the CSM that may be present at the surface of an exploding star. At 50\,d after breakout (or explosion), $V_{\rm max}$ can be inferred from the H$\alpha$ line, either using the maximum extent in velocity of the absorption trough if present, and otherwise using the width of the broad H$\alpha$ emission (as observed for example in SN\,1998S;  \citealt{leonard_98S_00}). This point has already be made in \citet{HD19} but using NLTE time-dependent simulations with \cmfgen. The corresponding $M_{\rm CDS}$ depends on a number of factors but for a standard type II SN explosion model like m15mlt3, Fig.~\ref{fig_vmax_vs_mcds} provides a reasonable estimate.

\section{The impact of detached CSM}
\label{sect_det}

The simulations presented so far exhibit a sudden rise to a bolometric maximum within a fraction of a day up to about a month depending on the CSM density and extent (see top panel of Fig.~\ref{fig_phot} and bottom-right panel of Fig.~\ref{fig_glob_prop}). None of these rising light curves match the step-like rise observed in SN\,2020tlf \citep{wynn_20tlf_22}. Such ``higher-than-expected" premaximum luminosity has been observed in a number of transients (e.g., SN\,2006oz). In the context of interacting SNe, such a premaximum luminosity may arise if the CSM is detached from the surface of the exploding star  \citep{moriya_maeda_12}.

\begin{figure}
\centering
\includegraphics[width=\hsize]{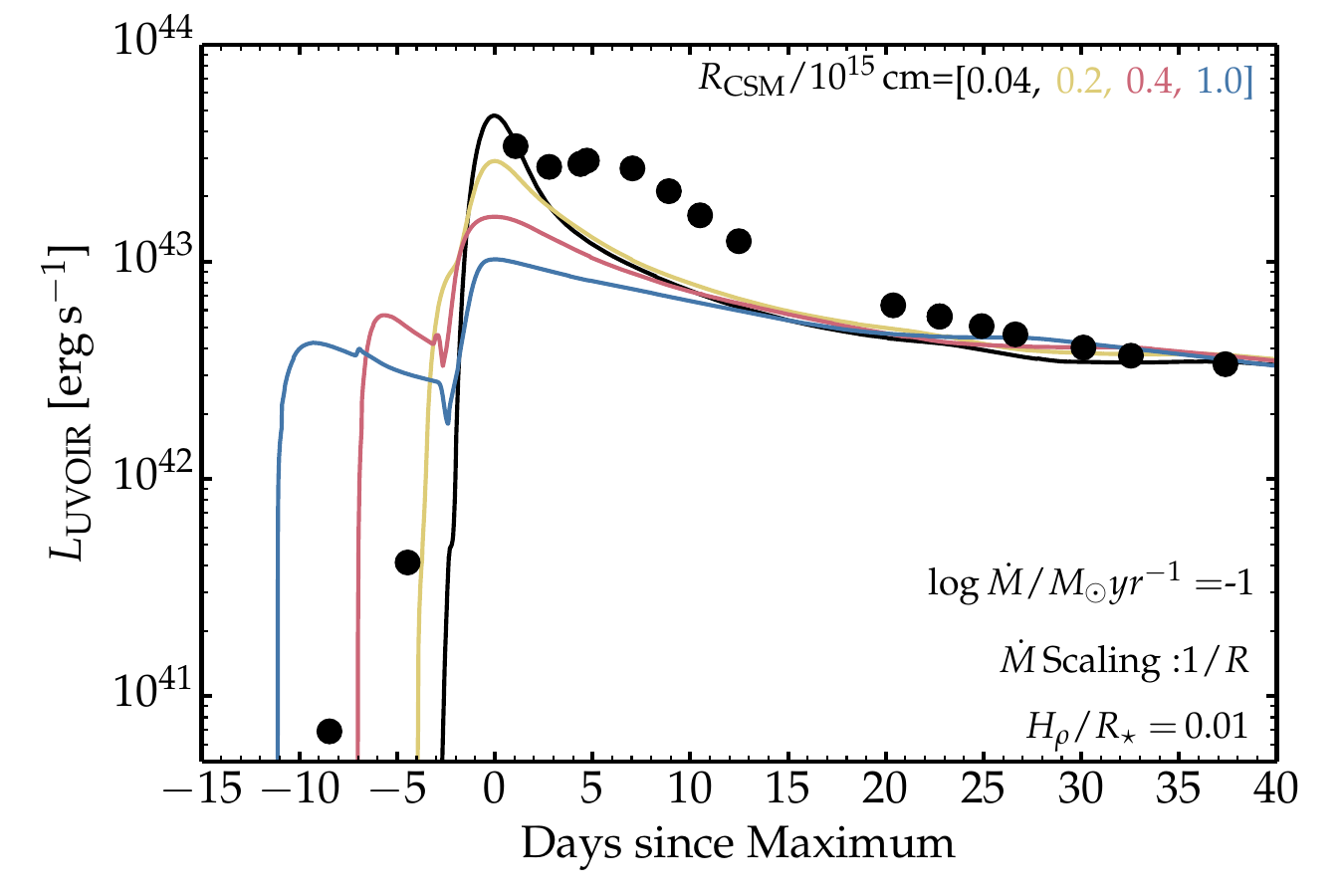}
\caption{UVOIR light curve for models in which the CSM is placed at 0.04, 0.2, 0.4, and $1 \times 10^{14}$\,cm above the stellar surface. Here, we show the model UVOIR luminosity in order to be more in line with the observations which include down to about 1500\,\AA\ (i.e., the bluest Swift filter; \citealt{wynn_20tlf_22}).
\label{fig_lbol_rcsm}
}
\end{figure}

In Fig.~\ref{fig_lbol_rcsm}, we show a set of simulations for initial ejecta-CSM configurations similar to case (1) but with a CSM that starts at a radius of 2, 4, and $10 \times 10^{14}$\,cm (an additional, undetached CSM model is shown for comparison). The corresponding initial density profiles are shown in Fig.~\ref{fig_init}.  We discuss only the case of a CSM wind density comparable to a wind mass-loss rate of 0.1\,\msunyr\ (with a more distant CSM, the wind density is smaller due to the larger radii so the impact of the interaction is reduced overall).

With a detached CSM, a first phase of high luminosity occurs as the radiation from the underlying ejecta (the breakout burst followed by the strong luminosity of the young SN) is reprocessed by the distant CSM. The more distant and extended the CSM, the longer-lived the  premaximum high luminosity, with a timescale that corresponds to the shock crossing time of the region between $R_\star$ and the base of the CSM (type IIn-like signatures would be seen during that phase if the CSM is sufficiently optically-thick despite being detached and therefore more tenuous). As the ejecta eventually collide with the CSM, a burst of radiation occurs but spread over the CSM diffusion time, which is longer for higher wind density (i.e., smaller inner radii for the CSM). Detached CSMs struggle to boost significantly the luminosity of the transient because the density at large distances suffers spherical dilution.

Something of the kind may be occurring in SN\,2020tlf although the models do not yield a satisfactory match to the observed light curve. In practice, one could envision a CSM density that would be lower closer to $R_\star$ and increase outward before declining at larger distances (see, for example, \citet{chugai_20tlf_22}). The density structure of the CSM is probably complicated and offset from an exact and continuous $1/R^2$ (or $1/R^3$, as used here) dependence.

\section{Conclusion and discussion}
\label{sect_conc}

We have presented a grid of radiation-hydrodynamics and NLTE radiative transfer calculations with the codes \heracles\ and \cmfgen\ for a RSG star exploding in CSM. We adopted extended, compact, or detached CSM configurations and documented the properties of the resulting ejecta and SN radiation, as well as how various quantities correlate with each other. We paid particular attention to the early-time properties when type IIn-like spectral signatures may be seen since these features are the unambiguous evidence of ejecta interaction with optically-thick CSM. This extends the previous study of \citet{HD19} in which the focus was on the aftermaths (i.e., at times greater than 15\,d post breakout) of an interaction taking place at earlier times (say over a few days at most after shock breakout).

We find that CSM with very different properties (mass, density, or extent) can produce essentially the same bolometric light curve. CSM mass (or mass-loss rate) estimates based on light curve modeling are therefore not uniquely defined. By studying in detail the radiation hydrodynamics of the ejecta-CSM interaction, we find that high density compact CSM yields no long-lived radiative precursor since the shock breaks out at the outer edge of the CSM, that is the dense CSM merely shifts the stellar radius outward. Lacking this phase, there is never any unshocked CSM to reprocess the radiation from the shock and produce the narrow lines with symmetric, electron-scattering broadened wings typical of type II-P SNe like 2013fs \citep{yaron_13fs_17}. Such configurations yield a luminosity boost for 10--20\,d after breakout because of energy deposition in the CSM, whose density is lower than in the progenitor envelope, allowing a faster release of that energy. Paradoxically, there is no ejecta-CSM interaction in this case since the dense CSM merely extends the star -- the shock progresses adiabatically through this CSM as it did through the stellar interior. A similar light curve may be obtained with a lower density but more extended CSM in which the photon-mean free path is large enough to allow the formation of a radiative precursor. Furthermore, being extended, the shock is decelerated for days and kinetic energy is extracted. In this configuration, there is a proper ejecta-CSM interaction and the luminosity boost is then caused both by the radiative precursor (radiation leakage from the radiation-dominated material behind the shock) as well as by the extra power arising from ejecta deceleration by the CSM.

An interesting paradox arises here. \citet{morozova_2l_2p_17}, \citet{morozova_sn2p_18}, or \citet{kozyreva_21yja_22} argue for a large CSM mass of 0.5\,\msun\ (and an associated mass-loss rate of order 0.1\,\msunyr) in objects that have essentially standard type II-P SN luminosity of a few 10$^{42}$\,\ergs.\footnote{This luminosity is small compared to the actual shock power. Indeed, for a SN shock ramming at velocity $V_{\rm sh}$ into a steady-state wind with mass-loss rate $\dot{M}$ and velocity $V_\infty$, the instantaneous power released by the interaction is
$L_{\rm sh} = \dot{M} V_{\rm sh}^3 / 2 V_\infty
= 3.15 \times 10^{40} \,\,\, \dot{M}_{-5} V_{\rm sh,4}^3 / V_{\infty,2} \,\,\, {\rm erg}\,{\rm s}^{-1} \, ,
$
where $\dot{M}_{-5} \equiv \dot{M} / 10^{-5} M_\odot$ yr$^{-1}$, $V_{\rm sh,4} \equiv V_{\rm sh}$/10000\,km\,s$^{-1}$ and $V_{\infty,2} \equiv V_\infty/100$\,km\,s$^{-1}$. With a mass-loss rate of 0.1\,\msunyr, the power released at the shock is of a few 10$^{44}$\,\ergs\ and rivals that inferred for superluminous SNe, though only if the radiation can escape.} The luminosity boost is small in those simulations because the density is so high that radiation cannot escape and therefore turns eventually mostly into kinetic energy.  By invoking large CSM masses and densities, the impact of the CSM is weakened rather than enhanced. Instead, a lower (but still high) CSM density is much more efficient at boosting the luminosity because the enhanced photon mean free path allows radiation to escape. It is this radiative leakage that boosts the SN luminosity and saps shock power. For example, a CSM mass of 0.4\,\msun\ was used to explain the observations of SN\,1998S \citep{D16_2n}, which is much more luminous than SN\,1999em for which \citet{morozova_sn2p_18} propose a CSM mass of 0.5\,\msun. In their modeling of SN\,1999em, they also argue for an ejecta kinetic energy of only $0.4 \times 10^{51}$\,erg, hence typical of underenergetic type II-P SNe \citep{lisakov_08bk_17}. Such a low ejecta kinetic energy is incompatible with the high velocity of the outer ejecta, as evidenced from the absorption at high velocity observed in H$\alpha$ or the Ca\two\ near-infrared triplet \citep{leonard_99em,DH06}, and indeed a standard-energy explosion with some modest amount of CSM is capable of reproducing the observations of SN\,1999em \citep{HD19}. Both photometric and spectroscopic constraints need to be matched by any adequate model.

Furthermore, we argue that the presence of type IIn-like signatures, as observed in SN\,2013fs or SN\,2020tlf \citep{wynn_20tlf_22} requires a high CSM density, but not too high so that a radiative precursor may form and allow the reprocessing of shock power by an extended optically-thick unshocked CSM. We find that type IIn signatures last until the CSM optical depth above the shock is about unity. As shown in the simulations of \citet{d17_13fs}, the short-lived presence ($\lesssim$\,1\,d) of narrow type IIn-like lines of O\four, O\five, O\six, or N\five\ in SN\,2013fs indicates a less extended but hotter CSM than in SN\,2020tlf which exhibited narrow type IIn-like lines of He\two\ or N\three\ but for a longer time (about one week).

From our radiation hydrodynamics calculations, we documented how various quantities track the dynamics of the ejecta-CSM interaction. While IIn signatures last for as long as there is optically-thick unshocked CSM, we also find that the CSM has a critical impact on the maximum velocity of the ejecta and its kinetic energy. Energy extraction that boosts the luminosity also translates into the disappearance of the fastest material in the ejecta, with obvious observable manifestations. The swept-up CSM and decelerated ejecta accumulates in a dense shell fostering the formation of a strong, broad H$\alpha$ emission line after a few weeks or months, as observed in SN\,1998S \citep{leonard_98S_00,D16_2n}.

RSG stars may both be surrounded by high-density CSM close to $R_\star$, contributing some luminosity boost, as well as dense, but not too dense, CSM present on larger scales to allow the formation of type IIn-like signatures for hours or days. It is those spectral signatures that set the greatest constraint on the CSM properties, including not just mass, density, and extent, but also temperature, ionization, and composition.
When the duration of these type IIn-like signatures is too short, it may be missed and one witnesses only the luminosity boost from the higher density compact CSM at the RSG surface, as inferred for  SN\,2021yja \citep{hosseinzadeh_21yja_22,kozyreva_21yja_22,vasylyev_21yja_22}.

The adoption of a large outer density, whether it is that of the CSM or the surface layers of a stellar evolution model, is problematic both numerically and physically. Numerically, a large density is often adopted in Lagrangian codes because there is no need to accommodate for the outer space into which the ejecta will expand. That is the outer grid point will expand as dictated by the motion of the outermost shell, thereby stretching the radial grid, but the outermost layer will always be at the same Lagrangian mass. In a Eulerian code like \heracles, this choice of outer density is invalid since the outer grid point must be placed very far above the initial extent of the star and CSM. However, from a physics standpoint, adopting a large outer density is inadequate because the entire grid becomes optically-thick at the time of shock breakout. In other words, the radiation cannot escape into optically-thin regions and the results for the shock breakout phase are compromised. This impacts the SN luminosity, but also the velocity, temperature, and density of the corresponding layers. To circumvent this issue, it is important to extend the RSG progenitor to low densities, whose original surface density is typically at $10^{-10}-10^{-9}$\,\gcc\ and therefore much too high (see, for example, \citet{DLW10a} for the RSG models of \citet{whw02}). This occurs because the gridding is set to handle the cool, recombined RSG atmospheres. When heated by the shock at breakout, these layers are immediately ionized and become optically thick, and their high density becomes immediately inadequate for radiation transport. While realistic 3D RSG models are the way forward \citep{goldberg_sbo_22}, one may in 1D, and for a start, extend the RSG atmospheres out to low density to allow for a proper physical handling of the breakout phase.

\citet{davies_csm_22} argued that a decade-long increase in mass loss at the RSG surface could lead to a dramatic and prolonged extinction of the underlying star until core collapse. Because such an extinction has so far not been observed, these authors argue that the CSM would need to form in a week-to-month outburst immediately before core collapse . This conclusion does not, however, rest on a physically-consistent dynamical model of the phenomenon. For a fixed energy supply, raising the mass-loss rate should deplete the radiative energy budget. But driving that mass loss in the first place will most likely occur through a significant boost to the energy supply in the RSG envelope and surface.  For example, in the extreme case of SN\,2020tlf, the super-wind phase that lasted for 120\,d prior to core collapse was associated not with a dimming of the progenitor but with its brightening, with a luminosity typical of a very massive star (much greater than the 12\,\msun\ progenitor mass inferred from nebular-phase spectra), and essentially at a constant effective temperature of about 4000\,K typical of a RSG star \citep{wynn_20tlf_22}. In general, outbursting stars brighten rather than fade, that is there is abundant supply of energy to both lift material out of the potential well and boost the luminosity of the event -- SNe are a common, albeit extreme example of this.

\begin{acknowledgements}

This research was supported by the Munich Institute for Astro-, Particle and BioPhysics (MIAPbP) which is funded by the Deutsche Forschungsgemeinschaft (DFG, German Research Foundation) under Germany's Excellence Strategy -- EXC--2094 -- 390783311. This work was supported by the ``Programme National de Physique Stellaire'' of CNRS/INSU co-funded by CEA and CNES. This work was granted access to the HPC resources of  CINES under the allocation 2020 -- A0090410554 and of TGCC under the allocation 2021 -- A0110410554 made by GENCI, France. This research has made use of NASA's Astrophysics Data System Bibliographic Services.

\end{acknowledgements}


\appendix

\section{Additional illustrations}

\begin{figure}
\centering
\includegraphics[width=\hsize]{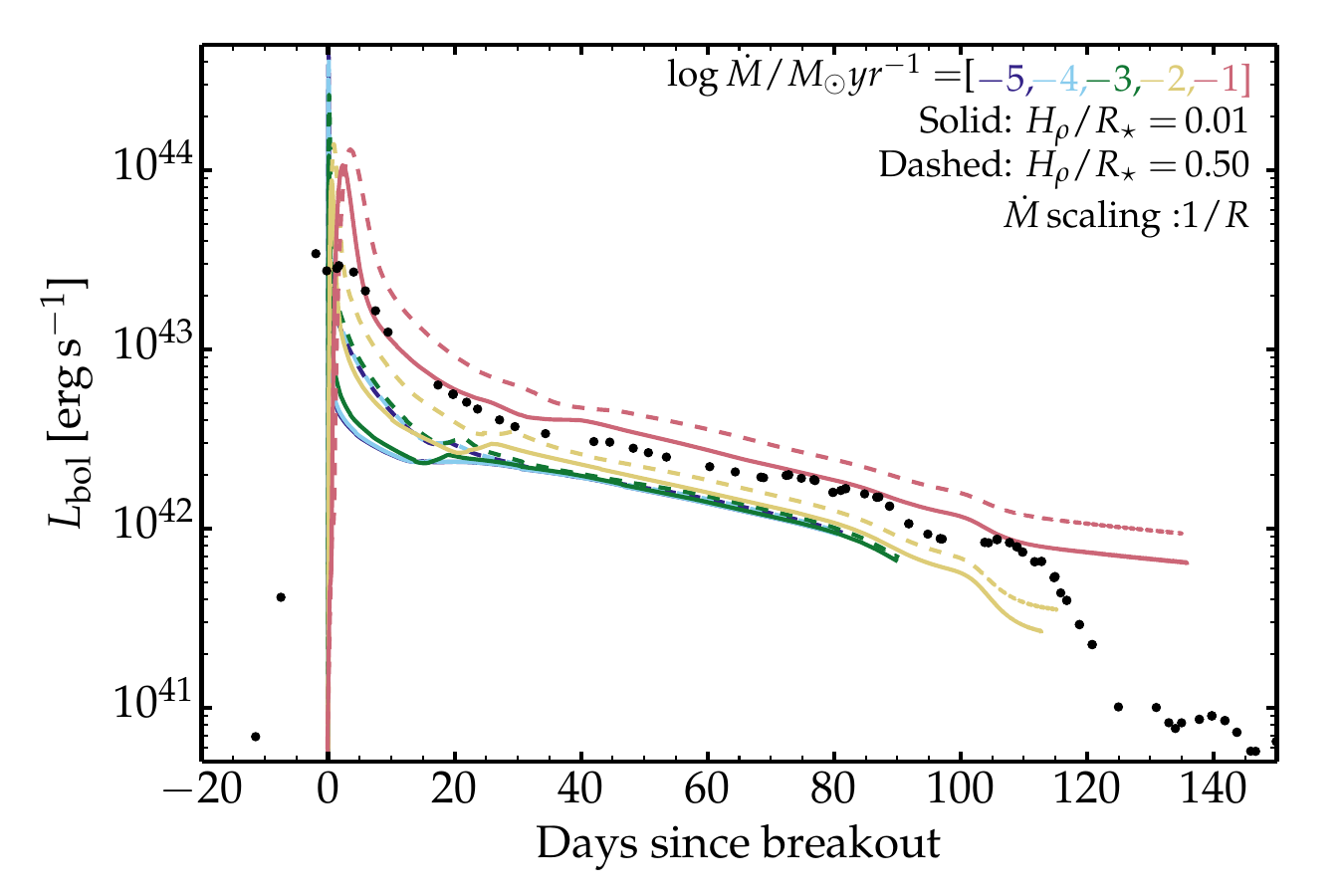}
\caption{\heracles\ light curves for different ejecta-CSM configurations. We compare the bolometric light curves obtained with \heracles\ for ejecta-CSM configurations corresponding to case-(1) and case-(2), as described in Section~\ref{sect_init}. We limit the sample for wind mass-loss rates covering from 10$^{-5}$ to 0.1\,\msunyr\ with an additional density scaling of $1/R$. The light curve degeneracy discussed in Sect.~\ref{sect_degen} is present throughout the sampled mass-loss rate range.
\label{fig_lbol_hd5em1_hd1em2}
}
\end{figure}

\end{document}